\documentclass[acmsmall,screen]{acmart}

\AtBeginDocument{%
  \providecommand\BibTeX{{%
    \normalfont B\kern-0.5em{\scshape i\kern-0.25em b}\kern-0.8em\TeX}}}

\setcopyright{acmcopyright}
\copyrightyear{2018}
\acmYear{2018}
\acmDOI{XXXXXXX.XXXXXXX}
\acmJournal{JACM}
\acmVolume{37}
\acmNumber{4}
\acmArticle{111}
\acmMonth{8}

\usepackage{tabularx}
\usepackage{subfigure}
\usepackage{graphicx}
\usepackage{multirow}
\usepackage{multicol}
\usepackage{bm}
\usepackage{setspace}
\usepackage{xcolor}

\DeclareMathOperator*{\argmax}{arg\,max}

\begin{document}

\title{An Analysis on Matching Mechanisms and Token Pruning for Late-interaction Models}

\author{Qi Liu}\authornote{Equal contribution.}
\email{liuqi\_67@ruc.edu.cn}
\orcid{0009-0000-4144-938X}
\affiliation{%
\department{Being Key Laboratory of Data Management and Analysis Methods, Gaoling School of Artificial Intelligence}
\institution{Renmin University of China}
\city{Beijing}
\country{China}
\streetaddress{No.59 Zhongguancun Street}
\postcode{100872}
}

\author{Gang Guo}\authornotemark[1]
\email{gguogangg@gmail.com}
\orcid{0009-0003-8516-3035}
\affiliation{%
\department{Being Key Laboratory of Data Management and Analysis Methods, Gaoling School of Artificial Intelligence}
\institution{Renmin University of China}
\city{Beijing}
\country{China}
\streetaddress{No.59 Zhongguancun Street}
\postcode{100872}
}

\author{Jiaxin Mao}\authornote{Corresponding author.}
\email{maojiaxin@gmail.com}
\orcid{0000-0002-9257-5498}
\affiliation{%
\department{Being Key Laboratory of Data Management and Analysis Methods, Gaoling School of Artificial Intelligence}
\institution{Renmin University of China}
\city{Beijing}
\country{China}
\streetaddress{No.59 Zhongguancun Street}
\postcode{100872}
}

\author{Zhicheng Dou}
\email{dou@ruc.edu.cn}
\orcid{0000-0002-9781-948X}
\affiliation{%
\department{Being Key Laboratory of Data Management and Analysis Methods, Gaoling School of Artificial Intelligence}
\institution{Renmin University of China}
\city{Beijing}
\country{China}
\streetaddress{No.59 Zhongguancun Street}
\postcode{100872}
}

\author{Ji-Rong Wen}
\email{jrwen@ruc.edu.cn}
\orcid{0000-0002-9777-9676}
\affiliation{%
\department{Being Key Laboratory of Data Management and Analysis Methods, Gaoling School of Artificial Intelligence}
\institution{Renmin University of China}
\city{Beijing}
\country{China}
\streetaddress{No.59 Zhongguancun Street}
\postcode{100872}
}

\author{Hao Jiang}
\email{jianghao66@huawei.com}
\orcid{0000-0002-2306-7596}
\affiliation{
\department{Distributed and Parallel Software Lab}
\institution{Huawei}
\city{Hangzhou}
\state{Zhejiang}
\country{China}
\streetaddress{No. 360 Jiangshu Road}
\postcode{310056}
}
\author{Xinyu Zhang}
\email{zhangxinyu35@huawei.com}
\orcid{0000-0002-6829-4522}
\affiliation{
\department{Distributed and Parallel Software Lab}
\institution{Huawei}
\city{Hangzhou}
\state{Zhejiang}
\country{China}
\streetaddress{No. 360 Jiangshu Road}
\postcode{310056}
}
\author{Zhao Cao}
\email{caozhao1@huawei.com}
\orcid{0000-0002-4214-7858}
\affiliation{
\department{Distributed and Parallel Software Lab}
\institution{Huawei}
\city{Hangzhou}
\state{Zhejiang}
\country{China}
\streetaddress{No. 360 Jiangshu Road}
\postcode{310056}
}

\renewcommand{\shortauthors}{Liu, et al.}

\begin{abstract}
  With the development of pre-trained language models, the dense retrieval models have become promising alternatives to the traditional retrieval models that rely on exact match and sparse bag-of-words representations. Different from most dense retrieval models using a bi-encoder to encode each query or document into a dense vector, the recently proposed late-interaction multi-vector models (i.e., ColBERT and COIL) achieve state-of-the-art retrieval effectiveness by using all token embeddings to represent documents and queries and modeling their relevance with a sum-of-max operation. However, these fine-grained representations may cause unacceptable storage overhead for practical search systems. In this study, we systematically analyze the matching mechanism of these late-interaction models and show that the sum-of-max operation heavily relies on the co-occurrence signals and some important words in the document. Based on these findings, we then propose several simple document pruning methods to reduce the storage overhead and compare the effectiveness of different pruning methods on different late-interaction models. We also leverage query pruning methods to further reduce the retrieval latency. We conduct extensive experiments on both in-domain and out-domain datasets and show that some of the used pruning methods can significantly improve the efficiency of these late-interaction models without substantially hurting their retrieval effectiveness.
\end{abstract}



\maketitle

\section{Introduction}\label{sec:intro}

Document ranking has always been one of the most important tasks in information retrieval. This task aims to rank documents according to their relevance to a given query. To achieve a good performance in terms of both effectiveness and efficiency, document ranking consists of two stages. The first stage is to retrieve thousands of documents to form a candidate set from a large corpus which may consist of millions or even billions of documents efficiently. Then in the second stage, the system usually uses a more sophisticated ranking model to re-rank the retrieved candidates to further boost the retrieval effectiveness.

Traditional Bag-of-Words (BoW) retrieval models (e.g., BM25~\cite{robertson2009probabilistic}) rely on sparse representations and exact matches between a query and documents to rank documents. While these methods are efficient, robust, and interpretable, the inverted index and the dependence on exact match signals may lead to the vocabulary mismatch problem~\cite{furnas1987vocabulary} which may limit the overall retrieval effectiveness. 

Recently, deep learning and pre-trained language models (PLMs) such as BERT~\cite{devlin2018bert}, RoBERTa~\cite{liu2019roberta} and~\cite{radford2019language, yang2019xlnet, lan2019albert} have made impressive improvements in many fields. In information retrieval, researchers have also leveraged the contextualized representation learned by PLMs to tackle the vocabulary mismatch problem and improve retrieval effectiveness. Some recent studies~\cite{dai2019context, dai2020context, bai2020sparterm, mallia2021learning, formal2021splade} leverage neural models to expand terms for original documents by modifying the term frequency or directly re-computing the term weight used in the traditional retrieval models. Other works~\cite{zhan2020repbert, karpukhin2020dense, luan2020sparse, khattab2020colbert, gao2021coil}, known as dense retrieval, represent queries and documents as one or more dense vector(s) and solve the vocabulary mismatch problem by performing soft matching in the vector space. The dense retrieval models have become promising alternatives to the traditional retrieval models because by using the approximate nearest neighbor search (ANNs) algorithm, they can achieve a good trade-off between retrieval effectiveness and efficiency.

For BERT-based retrieval models, the \emph{cross-encoder} architecture~\cite{nogueira2019passage} shown in Figure~\ref{subfig:arch-cross} makes thorough interaction between all query tokens and all document tokens in every transformer layer. Although such a full-interaction architecture can achieve a good ranking performance, it imposes a high time cost which prevents it from being used as a first-stage retrieval model. To solve this problem, some studies~\cite{zhan2020repbert, xiong2020approximate, karpukhin2020dense} propose to use a \emph{bi-encoder} architecture to independently encode queries and documents into dense vectors, and the architecture is illustrated in Figure~\ref{subfig:arch-bi}. Because there is no interaction between the query tokens and document tokens, we can pre-compute the embeddings for documents. When doing retrieval, the model only needs to encode the query and then search for the matching documents with the help of ANNs algorithms (e.g., Faiss~\cite{johnson2019billion}), within an acceptable latency for the first stage retrieval. However ~\citet{luan2020sparse} have theoretically proved and empirically demonstrated that using multiple vectors to represent documents is more effective than a single-vector representation. Therefore, recent studies~\cite{humeau2019poly,khattab2020colbert, gao2021coil} employ multiple vectors to represent a document in the first-stage retrieval and enhance the interaction between queries and documents. This \emph{late-interaction} architecture is shown in Figure~\ref{subfig:arch-late}. Combining the advantages of cross-encoders and bi-encoders, ~\citet{khattab2020colbert} proposed a soft late-interaction model named ColBERT, and ~\citet{gao2021coil} proposed a hard late-interaction model named COIL. These two models are more effective than the bi-encoders and much more efficient than the cross-encoders so they can be used as first-stage retrieval models. 

Although the above late-interaction models achieve a good balance between retrieval effectiveness and efficiency, their matching mechanisms are underinvestigated, which may undermine their interpretability and robustness, especially when compared to the traditional BoW-based retrieval models. Moreover, late-interaction models require to use all token embeddings to represent documents, leading to a significant increase in storage requirements and a decrease in efficiency. Therefore, it's important to understand their matching mechanisms and investigate whether and how we can leverage such matching mechanisms to reduce storage overhead and enhance efficiency. Specifically, in this paper, we propose several heuristic pruning methods according to the analysis of the matching mechanism, and study the following research questions:
\begin{itemize}
    \item \textbf{RQ1:} How do the late-interaction models work and perform matching at the token level? 
    \item \textbf{RQ2:} What are the differences in performance among various models when applying pruning?
    \item \textbf{RQ3:} How do different pruning methods and pruning ratios affect the effectiveness and efficiency of retrieval models?
\end{itemize}

Regarding \textbf{RQ1}, in this study, we first investigate how late-interaction models compute the relevance score for the query-document pair to understand the matching process of late-interaction models. We find that the computation of the relevance score heavily relies on the co-occurrence signals and the important words in documents. Such behaviors are very similar to the traditional BoW-based retrieval models in which we can safely ignore words that provide little information for relevance estimation, and use some weighting scheme, such as IDF, to emphasize the important words in the documents. 

Inspired by the matching mechanisms of traditional retrieval models and the similarity between late-interaction models and traditional retrieval models, we further ask: \emph{can we achieve comparable effectiveness when only using the important words in documents to compute the relevance score?} Therefore, we propose and evaluate several simple document token pruning strategies, which prune document tokens based on their positions, IDF values, and attention scores, to improve storage efficiency. Furthermore, we study query token pruning methods that reduce the query latency in the retrieval stage.

To address \textbf{RQ2} and \textbf{RQ3}, we conduct extensive experiments on a popular ad-hoc search benchmark (i.e., the MS MARCO dataset~\cite{bajaj2016msmarco}) and a popular zero-shot retrieval dataset (i.e., BEIR~\cite{thakur2021beir}) for different late-interaction models and different pruning strategies. Subsequently, we provide a thorough analysis of the results.

We believe the investigation of the matching mechanism can help us better understand these late-interaction models and the exploration of token pruning strategies is beneficial for reducing the storage overhead and query latency with minimal performance loss. In summary, our contributions are as follows:
\begin{itemize}
    \item To dig into the matching process of late-interaction models, we systematically analyze how ColBERT and COIL rank the documents based on the sum-of-max scoring function and find that it heavily depends on the informative words in documents. 
    \item We propose several simple but effective document token pruning methods and query token pruning methods for late-interaction models to improve efficiency.
    \item We conduct extensive experiments on both in-domain and out-of-domain datasets to compare the effectiveness of the different pruning methods on different neural retrieval models.
\end{itemize}

The rest of this paper is organized as follows. Section~\ref{sec:related_work} briefly reviews the related work. Section~\ref{sec:method} gives a detailed analysis of the late-interaction models and further proposes several methods of document tokens pruning and query tokens pruning. Section~\ref{sec:experiment} and Section~\ref{sec:results} describe the experiment settings and the empirical results, respectively. Section~\ref{sec:discussion} discusses the main findings in this paper and significant implications. Finally, we conclude this paper and list some potential limitations and future work directions in Section~\ref{sec:conclusion}.
\section{Related Work}\label{sec:related_work}

We will briefly review the previous works related to retrieval models and pruning techniques in this section.

\subsection{Traditional Retrieval Models}

\textbf{Traditional retrieval models} are based on sparse representations and hard-crafted rules such as exact lexical match models (e.g., BM25~\cite{robertson2009probabilistic}) and probability retrieval models (e.g., Query Likelihood~\cite{ponte1998language}). Take BM25 as an example, to retrieve documents, this type of model computes the relevance score between the query and document based on the co-occurrence terms. Benefiting from the inverted index, these models can retrieve documents efficiently so they are widely used as the first-stage retrieval model in practice. Besides the efficiency, the matching mechanism of these traditional retrieval models is transparent so that the researchers and engineers can understand how these models work. However, it is difficult for traditional retrieval models to handle the vocabulary mismatch problem which may limit their ability in modeling the semantic relevance between queries and documents. 

\subsection{Neural Retrieval Models} 

Recently, neural network models have attracted much attention in both academia and industry and have made impressive improvements in many fields. The information retrieval field benefits from pre-trained language models such as BERT ~\cite{devlin2018bert} and RoBERTa ~\cite{liu2019roberta}. These neural retrieval models can be divided into two groups as below:

\textbf{Neural Sparse Retrieval Models} improve retrieval performance by changing the term importance in different ways. Two common methods are: a) leveraging neural language generation models (e.g., T5~\cite{raffel2019exploring}) to expand original documents with semantic-related words, or b) using pre-trained language models to re-compute the weights. ~\citet{nogueira2019document,nogueira2019doc2query} modified the term frequency by using generative models to generate pseudo-queries of original documents and concatenate these pseudo-queries and original documents to form the new documents. ~\citet{dai2019context, dai2020context} used the pre-trained language model to predict contextualized term weight through a regression task. ~\citet{mallia2021learning} first used a generator to expand new semantic-related words of the original text and then used a contextualized language model encoder to re-compute term weights. On the other hand, ~\citet{bai2020sparterm} proposed using an important predictor to re-predict term weights and a gating controller to filter or expand terms at the same time. \citet{formal2021splade, formal2021spladev2} further proposed a more simple but effective method, based on explicit sparsity regularization and a log-saturation effect on term weights. All of these works can significantly outperform BM25.

\textbf{Neural Dense Retrieval Models} use dense vectors to represent queries and documents. Two widely adopted architectures of dense retrieval models are cross-encoders and bi-encoders. Cross-encoders~\cite{nogueira2019passage} concatenate the query and the document as the input of pre-trained language models, then use a function to map the output vector to a scalar score. Cross-encoders achieve superior retrieval effectiveness through the full interaction between all query tokens and all document tokens but the requirement of concatenating the query and each candidate document at the inference time makes cross-encoders impractical in the retrieval stage. 

Bi-encoders encode queries and documents independently so we can pre-compute the document representations (i.e., dense vectors), and then directly use these vectors for brute force search or ANNs in the retrieval phase. Bi-encoders can be further classified into two categories: the single-vector models and the multi-vector models. Single-vector models such as RepBERT~\cite{zhan2020repbert} and DPR ~\cite{karpukhin2020dense} use a single-vector to represent a document, while ~\citet{luan2020sparse} theoretically proved and empirically demonstrated that using multi-vectors to represent documents can achieve better performance. The recently proposed late-interaction models (e.g., ColBERT ~\cite{khattab2020colbert, santhanam2022colbertv2} and COIL ~\cite{gao2021coil}) achieve a great balance between effectiveness and efficiency by using all token embeddings to represent documents and queries, and modeling their relevance using a sum-of-max score. Although these late-interaction models achieved comparable effectiveness with cross-encoders, the unclear matching process and overhead storage are still problems that need to be solved.

\subsection{Pruning for Inverted Index}

Many studies have investigated the pruning of the traditional inverted index~\cite{carmel2001static, baeza2008design, buttcher2006document, blanco2007static, blanco2010probabilistic}. According to the different components of an inverted index that be pruned, they can be classified into different types. Some early approaches proposed to completely remove unimportant terms from the inverted index, according to relevance score~\cite{carmel2001static} or IDF scores~\cite{blanco2007static}. Further, researchers explored pruning methods based on term posting lists or documents, named term-centric pruning and document-centric pruning, respectively. In term-centric methods, the posting lists are pruned a proportion of postings based on the posting scores~\cite{carmel2001static, blanco2010probabilistic}. However, within document-centric approaches, the decision to prune is based on the rank of a posting within the corresponding document~\cite{thota2011within, buttcher2006document, altingovde2012static}. Meanwhile, ~\citet{altin2020pre} explored pruning the terms in documents before building the full index. Recently, ~\citet{lassance2023static} analyzed the pruning techniques for the learned term impact by neural sparse retrieval models, such as DeepImpact~\cite{mallia2021learning} and SPLADE~\cite{formal2021spladev2}.

Our proposed document token pruning methods can be considered as a document-centric approach. Compared to the above work, our research primarily focuses on the pruning of dense vector indexed composed of token embeddings.

\subsection{Analysis and Pruning for (Col)BERT}

For BERT in information retrieval, \citet{qiao2019understanding} and \citet{zhan2020analysis} analyzed the behaviors of BERT when used as a cross-encoder for ranking. For ColBERT, researchers could analyze it at the token level as it uses the sum-of-max operation and some specific tokens will be activated during matching. ~\citet{formal2021white} analyzed ColBERT's matching behavior through term importance and exact/soft matching patterns by computing the correlation between the ranking results after masking different tokens and proved ColBERT does have a notion of token importance. Different from ~\cite{formal2021white}, we analyze both COIL and ColBERT in a more fine-grained perspective by investigating the contribution of different tokens to the final sum-of-max score.

Such analytical studies have provided valuable heuristic insights for pruning ColBERT. ~\citet{tonellotto2021query} computed the IDF values of all documents and only used a few fixed query token embeddings according to the IDF values to retrieve fewer documents. The number of retrieved documents is much smaller than the original and experimental results show that query pruning can speed up retrieval of ColBERT without significant performance loss. Moreover, ~\citet{lassance2021study} explored using several documents token pruning methods on ColBERT, results suggested that ColBERT indexes can be pruned up to 30\% on the MSMARCO passage collection without a significant drop in performance. However, \citet{lassance2021study} selected a fixed number of tokens for different-length passages, which is not reasonable, as long passages tend to have more complicated meanings. Different from ~\cite{lassance2021study}, we use pruning methods in a dynamic way.

Additionally, some studies employed different methods to reduce the storage of ColBERT at training or indexing time. ~\citet{hofstatter2022introducing} reduced the number of vectors by learning unique whole-word representations and using a learned gate vector. \citet{qian2022multi} employed entropy-regularized linear programming to achieve the sparsity of the matching matrix. On the other hand, ~\citet{santhanam2022colbertv2} used an aggressive residual compression mechanism to improve the index storage. While these approaches require complex training and indexing techniques, our work focuses on simply pruning tokens without training and demonstrates that such pruning strategies are simple yet effective.

In addition to existing studies, we systematically analyze both COIL and ColBERT in a more fine-grained method, which investigates the scores of different tokens contributing to the final sum-of-max score. Moreover, we conduct extensive experiments which include several different token pruning methods on both COIL and ColBERT, as well as some sparse retrieval models.

\section{Methodologies}\label{sec:method}

In this section, we first provide a brief overview of the document retrieval task, as well as a discussion of three prevalent architectures of dense retrieval models utilized in this task. Then we conduct a detailed analysis of the late-interaction architecture, indicating how the late-interaction mechanism works. Building upon our analytical findings, we propose several methods of document token pruning and query token pruning.

\subsection{Preliminaries}

\begin{figure}[t]
    \centering
    \subfigure[cross-encoders]{
        \begin{minipage}[t]{0.5\linewidth}
            \centering
            \includegraphics[width=0.8\textwidth]{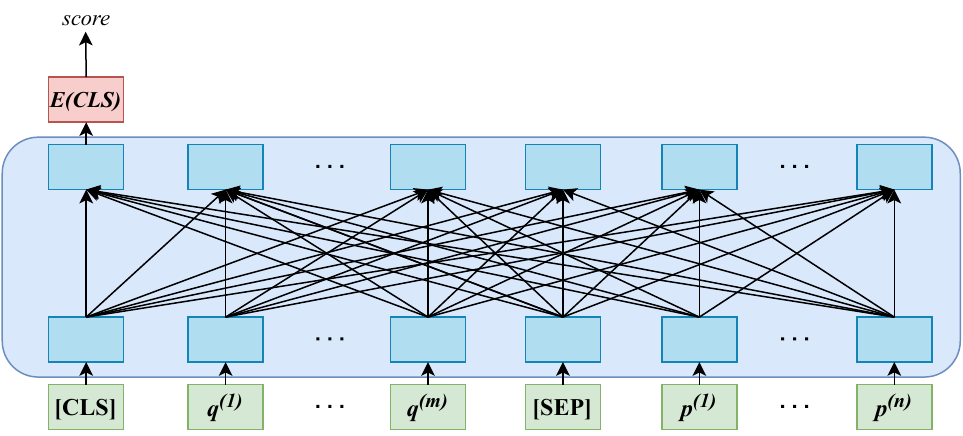}
            \label{subfig:arch-cross}
        \end{minipage}%
    }%
    \subfigure[bi-encoders]{
        \begin{minipage}[t]{0.5\linewidth}
            \centering
            \includegraphics[width=0.8\linewidth]{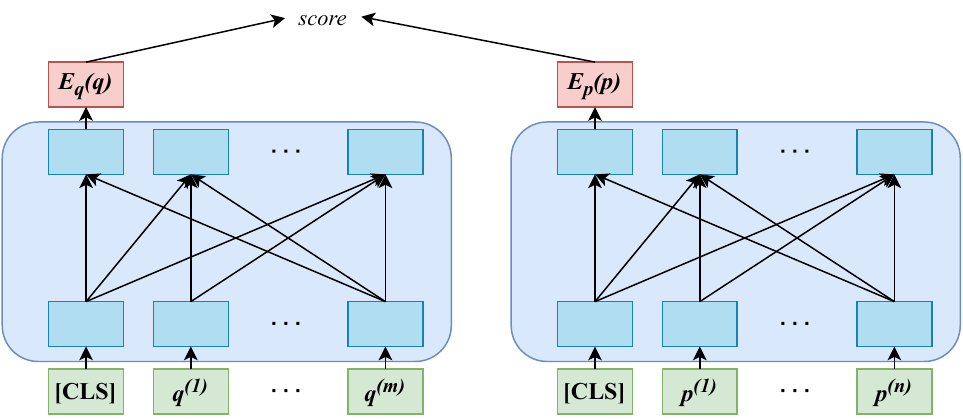}
            \label{subfig:arch-bi}
        \end{minipage}%
    }%
    \\
    \subfigure[late-interaction models]{
        \begin{minipage}[t]{0.5\linewidth}
            \centering
            \includegraphics[width=0.8\linewidth]{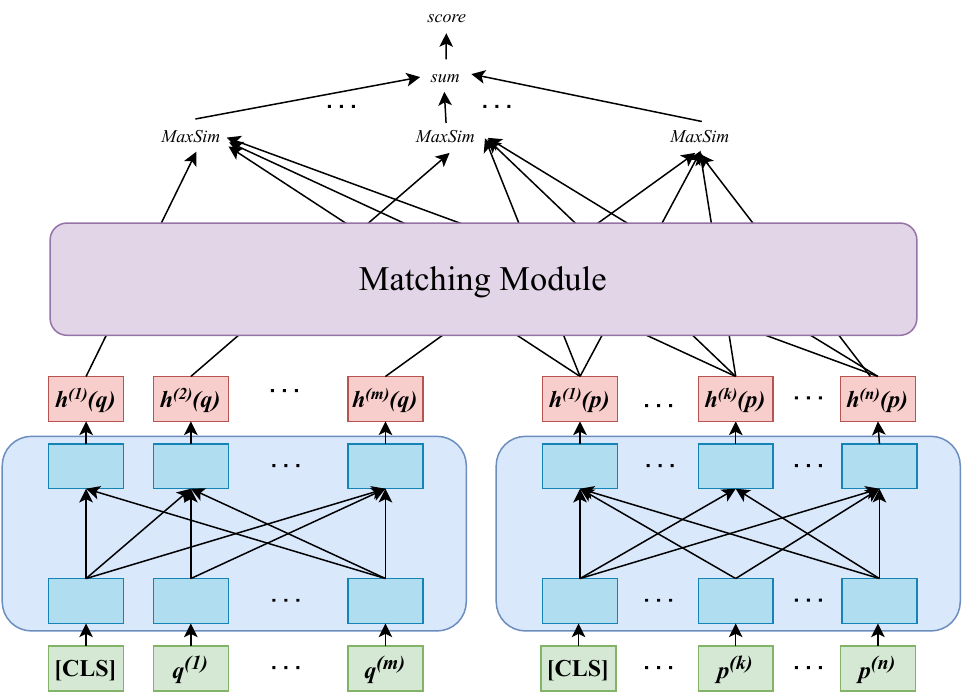}
            \label{subfig:arch-late}
        \end{minipage}%
    }%
    \caption{Architectures of dense retrieval models.}
    \label{fig:architectures}
\end{figure}

For the document retrieval task, given a query $q$ and a large-scale document collection $\mathcal{D}$, we need to retrieve a set of documents that are most relevant to the query $q$. The core problem here is how to compute the relevance score, denoted as $s(q, d)$, between the given query $q$ and the document $d \in \mathcal{D}$. In recent years, researchers have employed PLMs to effectively model the relevance between queries and documents. Consequently, they have proposed several different architectures for this purpose. Here, we will briefly introduce and give the formal definition of the three distinct architectures, which are shown in Figure~\ref{fig:architectures}.

The \emph{cross-encoders} (Figure~\ref{subfig:arch-cross}) take the concatenation of the query tokens and the document tokens as input so that the models can compute the attention scores between all query tokens and all document tokens in every transformer layer. This architecture allows rich interaction between the query and the document, which leads to a great retrieval performance. On top of the transformer, there is a function, usually a linear projection, to map the output embedding, such as the average embeddings of all tokens or the [CLS] embedding, to a scalar value which is regarded as the relevance score of the given query and the document~\cite{nogueira2019passage}.

The \emph{bi-encoders} (Figure~\ref{subfig:arch-bi}) encode queries and documents independently and then reduce the output of all token embeddings of the query to a single embedding and also for the document. The input is the independent query and document. While there are two encoders in this logical architecture, in practice, some use two different encoders, while others use a parameter-shared one. After the query and document are encoded by the encoder(s), we can get their corresponding token embeddings, and use the embedding of the [CLS] token \cite{karpukhin2020dense} or mean values of all tokens embeddings \cite{zhan2020repbert} as the representation of the whole query or document. The inner product or cosine similarity between the query embedding and the document embedding is usually used as the final relevance score. Compared to cross-encoders, as the queries and documents can be computed independently, bi-encoders achieve efficient retrieval by encoding all documents offline and only using one embedding to represent queries and documents. Additionally, it allows the use of the approximate nearest neighborhood search (ANNs) techniques to further reduce the online retrieval latency. However, the heavy reduction of token embeddings may result in a great loss of information and limit the retrieval effectiveness.

The \emph{late-interaction models} (Figure~\ref{subfig:arch-late}) combine the advantages of the previous two methods, and independently encode the query and the document at the token level. It employs a matching module referred to as ``late interaction'' to compute the relevance score between the query and the document, which makes a good trade-off between efficiency and effectiveness as it can achieve comparable effectiveness as cross-encoders and comparable efficiency as bi-encoders. Most of the existing late-interaction models utilize a sum-of-max operation~\cite{khattab2020colbert} to compute the ranking score, which can be described as:
\begin{equation}
     Q = PLM(q) \in \mathbb{R}^{m \times h}, \quad 
     D = PLM(d) \in \mathbb{R}^{n \times h},
\end{equation}
\begin{equation}
s(q,d) = \textit{Match}(Q, D) = \sum_{q_{i} \in M_q} \max_{d_{j} \in M_d} Q_{i} \cdot D_{j}^T,
\label{sum-of-max}
\end{equation}
where $Q$ represents the matrix of all query token embeddings, $D$ represents the matrix of all document token embeddings, and $m$ and $n$ are the length of query and document, respectively. The two sets $M_q$ and $M_d$ represent the sets of query tokens and document tokens, respectively, which will be used in the matching. Equation~\eqref{sum-of-max} denotes the sum-of-max operation which is adopted in late-interaction models. 

According to how $M_q$ and $M_d$ are constructed, the matching module can be implemented in two ways: \emph{soft matching}, which aligns with ColBERT~\cite{khattab2020colbert}, or \emph{hard matching}, which aligns with COIL~\cite{gao2021coil}. For ColBERT, each query token embedding will interact with each document token embedding, regardless of if they are same. We regard this as the \emph{soft matching}:
\begin{equation}
    \textit{Soft-Match}(Q, D) = \sum_{i=1}^m \max_{j=1}^n Q_{i} \cdot D_{j}^T.
\end{equation}
As for COIL, only the token that occurs in both query and documents will be selected, and each query token only interacts with the same tokens in the document. As it performs like the traditional exact match, we regard this as the \emph{hard matching}:
\begin{equation}
    \textit{Hard-Match}(Q, D) = \sum_{q_i \in q \cap d} \max_{d_j = q_i} Q_{i} \cdot D_{j}^T.
\end{equation}
The documents encoder and queries encoder of ColBERT and COIL both share parameters so that they can map queries and documents into the same semantic space. 

In order to perform the sum-of-max operation, the two models pre-compute and store all the token embeddings of document tokens. This sum-of-max operation allows the interaction between all query tokens' embeddings and all document tokens' embeddings after the last transformer layer without embeddings reduction compression so that they can achieve better retrieval performance than bi-encoders. However, such late-interaction models store all document token embeddings through the interaction between all query tokens and all document tokens after the last transformer layer, which may result in a large storage overhead and a relatively long query latency, respectively. In this paper, we aim to tackle these two limitations with document token pruning and query token pruning strategies.

\subsection{Analyses of late-interaction models} \label{sec:matching}

To understand how the late-interaction mechanism works, we first analyze the matching mechanism of ColBERT and COIL when performing document ranking. In particular, as the final relevance score is a sum of several matching scores between the query tokens and the document tokens selected by the max operation, we investigate \emph{which} document tokens will be selected and \emph{how much} they contribute to the final relevance score.

For the first question, we contemplate the position and IDF value of each token within a document. Specifically, due to the disparate lengths of the documents, we partition the tokens into ten bins of equal size according to their relative position or IDF values. For example, in the case of position-based partitioning, the first bin encompasses the initial $10\%$ of tokens within the document, while in the case of IDF-based partitioning, the first bin comprises the tokens with 10\% IDF values in the document. Note that we exclude all special tokens, such as the [CLS] and [SEP] tokens, in this analysis.

After partitioning tokens into different bins, we can assess their contribution to the final score for each bin, corresponding to the second question. In particular, we define two contribution metrics for one document token, named \emph{indices contribution} and \emph{score contribution}. The indices contribution represents whether the token is selected by the max operation in Equation~\eqref{sum-of-max}, while the score contribution denotes the corresponding inner product score which contributes to the final score $s(q, d)$. Formally, we compute these two contribution metrics of each bin across a set of positive query-document pairs, and the whole process can be described as follows:
\begin{equation}
    P_{\textit{Indice}}(k) = \frac{1}{Z} \sum_{\{q, d\}} \sum_{q_i \in M_q} I \left(\argmax\limits_{d_j \in M_d} Q_i \cdot D_j^T \in \textit{Part}(d, k) \right),
\end{equation}
\begin{equation}
    P_{\textit{Score}}(k) = \frac{1}{Z} \sum_{\{q, d\}} \sum_{q_i \in M_q}  \left( \max_{d_j \in M_d} Q_i \cdot D_j^T \right) \cdot I \left(\argmax\limits_{d_j \in M_d} Q_i \cdot D_j^T \in \textit{Part}(d, k) \right)
\end{equation}
where $I(\text{condition})$ is an indicator function, i.e., $I(\text{condition}) = 1$ if the condition is true $0$ otherwise, $\textit{Part}(d, k)$ is the $k$-th bin of the document $d$ partitioned by tokens' relative position or IDF values, and $Z$ is the normalized factor.

For comparison, we include a hypothetical reference contribution in our experiment. If we assume that all tokens are contribute equally in the matching process, then for a sufficiently large set, each token within a bin should have the nearly same indices contribution and score contribution. In fact, both original ColBERT and COIL store the embeddings of all tokens. Therefore, we use the uniform contribution $P_{\textit{Hypo}}(k) = \frac{1}{10}$ as a reference.

\begin{figure}[t]
\centering
    \subfigure[Position Contribution of ColBERT]
    {
    \includegraphics[width=0.45\textwidth]{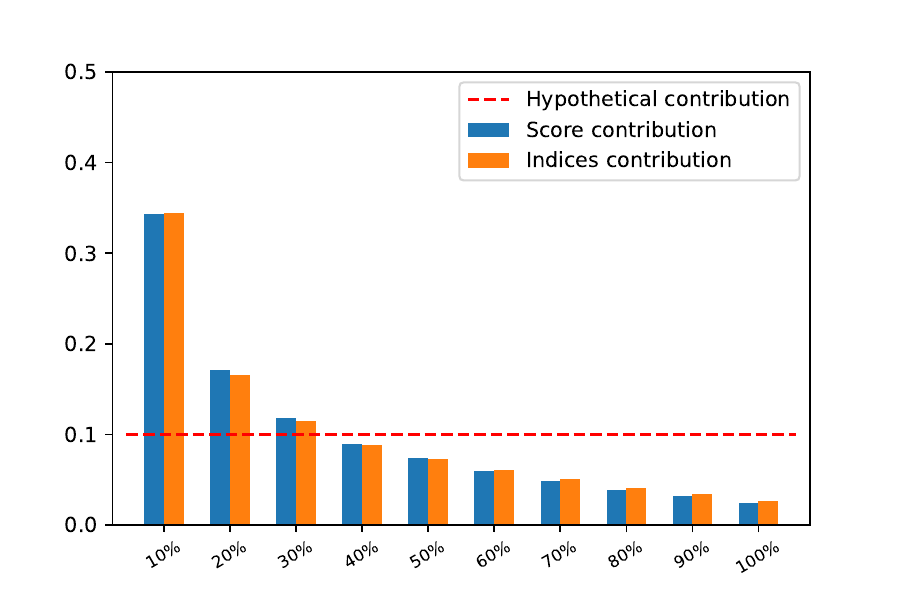}
    \label{subfig:contribution-colbert-position}
    }
    \subfigure[IDF Contribution of ColBERT]
    {
    \includegraphics[width=0.45\textwidth]{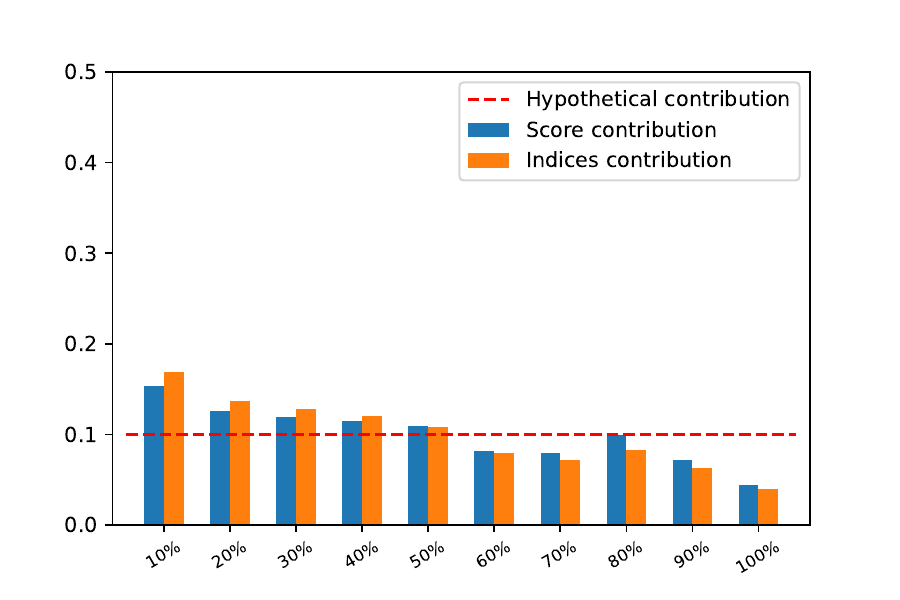}
    \label{subfig:contribution-colbert-idf}
    }
    \subfigure[Position Contribution of COIL]
    {
    \includegraphics[width=0.45\textwidth]{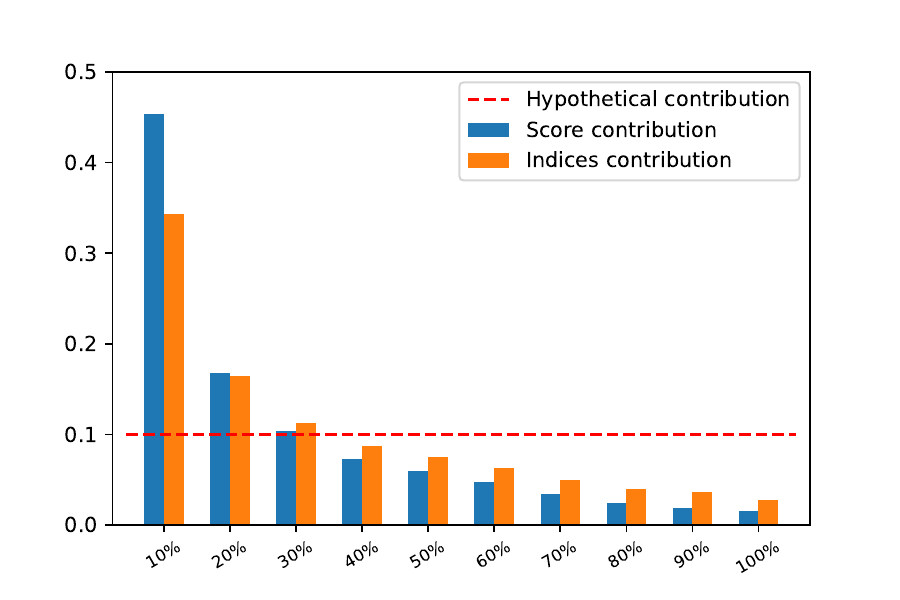}
    \label{subfig:contribution-coil-position}
    }
    \subfigure[IDF Contribution of COIL]
    {
    \includegraphics[width=0.45\textwidth]{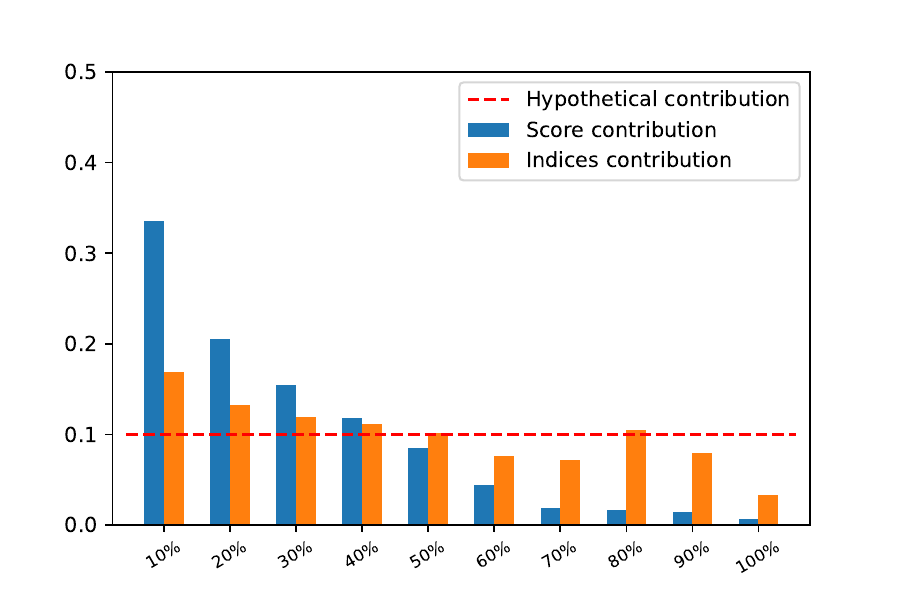}
    \label{subfig:contribution-coil-idf}
    }
\caption{The distribution of contributions of different tokens. Each bin consists of ten percent of the total tokens according to their relative position or IDF values. The y-axis shows the proportion contributing to the final scores of each bin.}
\label{fig:contribution}
\end{figure}

We empirically study how the document tokens with different positions and different IDF values contribute to the final relevance score and conduct the experiment on the training set of MS MARCO. The results are shown in Figure~\ref{fig:contribution}. In general, we can see that the tokens with different positions and IDF values contribute rather differently to the final relevance score which means that while the model is matching queries and documents, the priority of each token is different. Furthermore, the token contribution is almost negatively correlated with positions and IDF values, indicating that the words that have a higher IDF value or appear at the beginning of the documents are more important when computing the relevance score. Moreover, this phenomenon is more significant in COIL compared to ColBERT.

Look into ColBERT, when we do the position-matching analysis whose result is shown in Figure~\ref{subfig:contribution-colbert-position}, both indices and score contributions of the ten position bins are decreasing. We suggest that this is because authors get used to writing important words at the beginning of a document. Another perspective is IDF-matching, as we can see in Figure~\ref{subfig:contribution-colbert-idf}, there are five bins that contribute more than the hypothetical contribution, and tokens in the last five bins fail to reach the hypothetical contribution, this can confirm that, while ColBERT matching, words having a higher IDF value are more significant.

For COIL, the position-matching contribution, as shown in Figure~\ref{subfig:contribution-coil-position}, demonstrates that both score and indices contribution consistently decrease according to the position. There are three bins' contributions higher than the hypothetical contribution, and the hypothetical contribution is not reached by the last seven bins. The score contribution drops faster than the indices contribution which means the average score of individual tokens in every bin decreases. We can deduce from the two mentioned explanations that the preceding token is more essential. Although there are modest swings in indices contribution in the IDF-matching shown in Figure~\ref{subfig:contribution-coil-idf}, the tendency is that IDF-indices contribution decreases as IDF values decrease. Besides, the IDF-score contribution consistently decreases, and the indices contribution becomes smaller than the score contribution since the fifth bin. We can deduce from Figure~\ref{subfig:contribution-coil-idf} that tokens with low IDF values contribute less to the final score than tokens with high IDF values.

Comparing the above two systems, when looking into the Position Contribution, we can see that ColBERT has a narrower contribution range and smoother volatility than COIL. Specifically, the score contribution and indices contribution of ColBERT is from (34.3\%, 34.4\%) to (2.4\%, 2.6\%) and COIL is from (45.4\%, 34.2\%) to (1.6\%, 2.8\%). For the IDF contribution, ColBERT's discrepancy between score contribution and indices contribution is substantially smaller than COIL's due to the different matching processes of ColBERT and COIL (soft matching v.s. hard matching). Revisit the concept of soft matching and hard matching: soft matching in ColBERT will choose the highest-scoring token from all tokens, whereas in COIL, hard matching will choose only from the same tokens, so COIL may hold more similarities compared to traditional retrieval models.

Moreover, for ColBERT, we analyze how the tokens that occur in both the query and document as well as the stop words and non-stop words contribute to the relevance score. Table~\ref{tab:contribution} shows the score contribution of concurrence tokens. From the results, we can see that ColBERT allows the sum-of-max operation to interact with any document tokens, but the document tokens which also appear in the query still obtain much higher attention scores. Table~\ref{tab:contribution} also shows the contributions of stop words and non-stop words to the final relevance score. The results show that ColBERT also pays less attention to stop words just like traditional retrieval models.

In summary, these results demonstrate that the behavior of ColBERT and COIL is similar to the traditional retrieval models, that is, paying more attention to important words and exact match. Therefore, we further propose to prune some tokens for these late-interaction models.

\begin{table}[t]
  \caption{Contribution to the final relevance score. Co-occurrence means the contributions of score that the token appears in both the query and the document.}
  \label{tab:contribution}
  \begin{tabular}{cc|cc}
    \toprule
    Co-occurrence & Others & Non-stop words & Stop words  \\
    \hline
    $69.1\%$ & $30.9\%$ & $74.3\%$ & $25.7\%$  \\
  \bottomrule
\end{tabular}
\end{table}

\subsection{Document Token Pruning}

\begin{figure}[t]
\centering
\includegraphics[width=\textwidth]{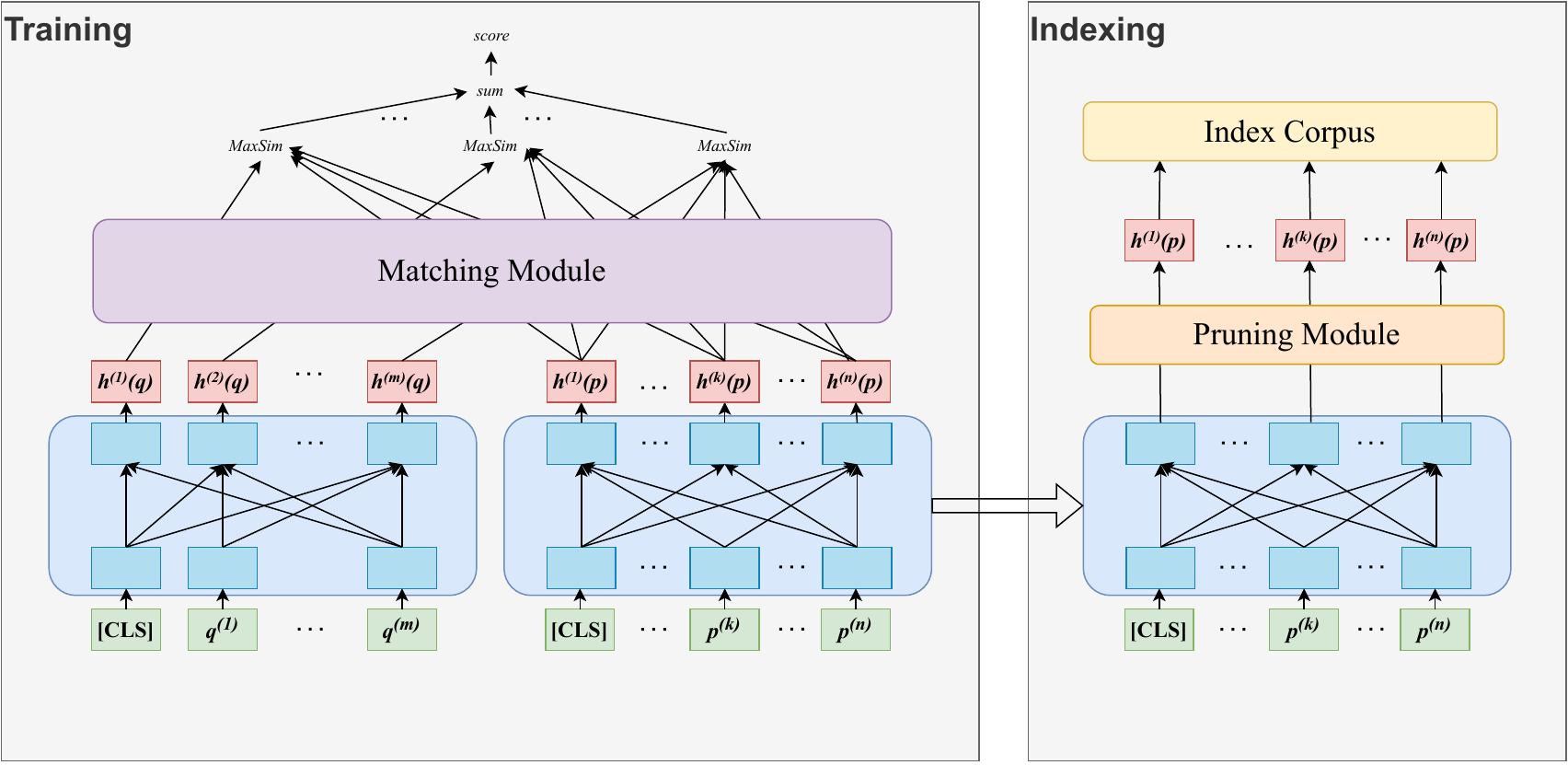}
\caption{The pipeline of Document Token Pruning}
\label{Fig:DTP}
\end{figure}

We describe the Document Token Pruning (DTP) pipeline in this section, shown in Figure~\ref{Fig:DTP}. We employ the pruning methods only during the process of inference and indexing, which we name post-pruning, rather than pruning at the input of the model or combining pruning with model training. This is because we have found that the post-pruning methods result in less performance loss than pruning at the input or simply adding a regularization during training in the pilot experiment. Moreover, the utilization of post-pruning can save a significant amount of resources because it avoids additional training of the pruned model. We train late interaction models following the original settings in~\cite{khattab2020colbert, gao2021coil}, and more training details can be found in Section~\ref{sec:implementation}.

After training models, we add a pruning module to determine whether to store the token embedding. Based on the analysis results presented in Section~\ref{sec:matching}, we propose three heuristic methods for pruning tokens to investigate how the performance of these late-interaction models changes with different pruning methods. All pruning methods perform in a dynamic way and tend to preserve a given portion of important tokens. More specifically, we will designate a \emph{remaining ratio}, denoted as $\alpha$ in the following, which determines the percentage of tokens from various documents that will be retained in the index. In other words, for each document, only $l' = \lfloor l \times \alpha \rfloor $ tokens' embeddings from it will be stored. Here $l$ is the length of the original tokenized document. In the following, we introduce the three pruning methods we propose in detail.

\paragraph{First-$\alpha$ Pruning.} The first pruning method entails retaining the initial $\alpha \times 100\%$ proportion of tokens in each document. We employ this approach due to the tendency of document authors to include crucial information at the beginning of the document. This method will be a simple but effective one.
\begin{equation}
    D' = D[1:l']
\end{equation}

\paragraph{IDF-Top-$\alpha$ Pruning.} Another method we propose is IDF-Top-$\alpha$ pruning, where document tokens are pruned based on their IDF values. IDF is a commonly employed term weighting schema in traditional retrieval models, which is both straightforward and highly effective. To implement this method, we arrange the tokens of documents in descending order of their IDF values, then retain and store the embeddings of the document tokens that correspond to the top $\alpha \times 100\%$ IDF value.
\begin{equation}
    \textit{Top-IDF-Index} = \textit{Top-}l' (IDF(d))
\end{equation}
\begin{equation}
    D' = D[\textit{Top-IDF-Index}]
\end{equation}

\paragraph{Attention-Top-$\alpha$ Pruning.} The third method is Attention-Top-$\alpha$ pruning. Attention score is a common metric to evaluate the importance of a token, here we take the interaction between document token embeddings as the self-attention matrix and take the column sum of the matrix as the importance of a token. Note that the self-attention matrix is symmetric, so the column sum and the row sum are the same. The Attention-Top-$\alpha$ pruning can be formalized as:
\begin{equation}
    A = \textit{softmax}(DD^T)\cdot 1_{n}
    \label{eq:att}
\end{equation}
\begin{equation}
    \textit{Top-Att-Index} = \textit{Top-}l' (A)
\end{equation}
\begin{equation}
    D' = D[\textit{Top-Att-Index}]
\end{equation}

Notice that we always retain the special tokens that appear at the beginning of the sequence, specifically [CLS] and [D] for ColBERT, and [CLS] for COIL. We do this because the embedding of special tokens is usually considered to encapsulate the information of the entire textual passage. Therefore, for the latter two approaches, the resulting index is, in fact, comprised of $l'$ tokens that include special tokens as well as tokens with the highest IDF values or attention scores.

To summarize, the document token pruning methods can be expressed as follows:
\begin{equation}
\begin{aligned}
   Q = PLM(q), \quad D' =& \ DTP(PLM(d)), \\
   \quad DTP \in \{ \text{First-}\alpha, \text{IDF-Top-}\alpha, & \ \text{Attention-Top-}\alpha \}
\end{aligned}
\end{equation}
\begin{equation}
    s(q, d) = \textit{Match}(Q, D').
\end{equation}

\subsection{End-to-End Retrieval and Query Token Pruning} \label{sec:QTP}

\begin{figure}[t]
\centering
    {
    \includegraphics[width=\textwidth]{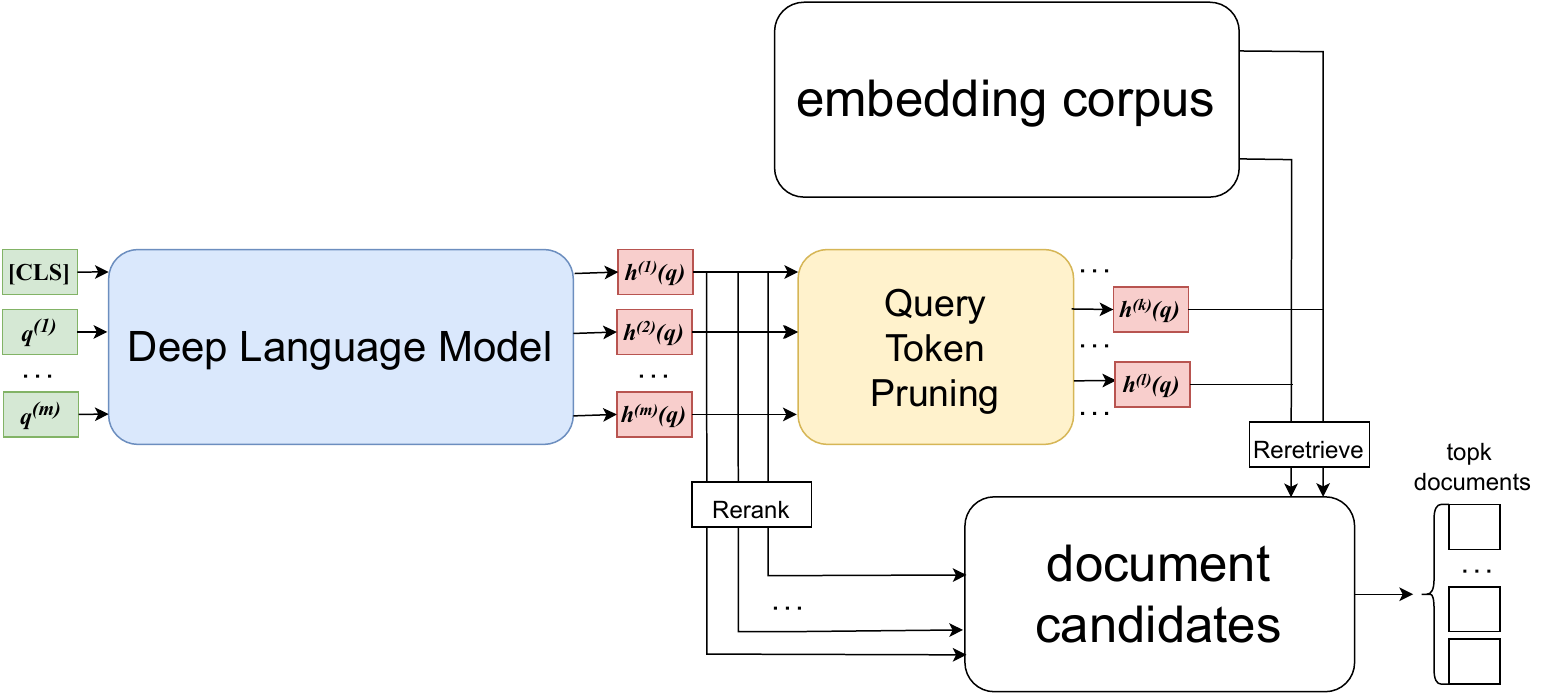}
    \label{QTP}
    }
\caption{The pipeline of Query Token Pruning for ColBERT.}
\label{Fig:QTP}
\end{figure}

We introduce how different late-interaction models perform end-to-end retrieval in this section. Specifically, for ColBERT, we conduct a detailed analysis to investigate how the query token retrieves relevant documents. Based on the empirical findings, we further propose Query Token Pruning (QTP) methods for ColBERT to reduce retrieval latency.

Due to the requirement of token-level interaction and the utilization of sum-of-max operation in the late interaction model, we can not directly retrieve the top-$k$ relevant documents from a dense vector index built by toolkits like Faiss~\cite{johnson2019billion} which only support inner product and L2 distance. Instead, we need to employ more complex pipelines.

ColBERT employs a two-stage retrieval pipeline: in the first stage it approximates the final relevance score and retrieves a set of candidates, then in the second stage the accurate score will be computed for ranking. In particular, in the first stage, ColBERT retrieves a fixed number of top document tokens from the Faiss index for every query token, then collects the corresponding documents to construct the candidates set. Then, we gather all document tokens for each document in the candidate set to perform a sum-of-max operation and compute the final score between the query and the documents for ranking. To reduce approximation error introduced by different score functions, it is typical to retrieve more documents.

COIL is based on an inverted index, it first calculates scores for each exact matched token in the query, then aggregates all matched tokens to produce a document score and uses heap sort to keep $k$ documents with the highest relevance score. While performing the sum-of-max operation, the inverted index can obtain an accurate score, so the inference stage and training stage can have the same score function.

As we discussed, COIL computes accurate scores while performing retrieval. However, ColBERT first approximates the relevance score and retrieves abundant documents for each query token, resulting in significant retrieval latency. Consequently, it is natural to explore the possibility of pruning certain query tokens for ColBERT to accumulate the retrieval process while minimizing any potential decrease in performance.

The pipeline of Query Token Pruning (QTP) is shown in Figure~\ref{Fig:QTP}. ColBERT performs Top-$k$ ANNs with every query token, so a large number of candidate documents will be retrieved. The model will spend a relatively long time transferring the embeddings of the candidate document from CPU to GPU to compute the sum-of-max ranking scores. However, ~\citet{tonellotto2021query} have shown that not every query embedding will bring useful documents for ranking. Therefore, performing Top-$k$ ANNs with every query token will impose unnecessary retrieval latency. ~\citet{tonellotto2021query} have proved that only query embeddings with the highest IDF values can be used for effective retrieval, which also confirms the contribution of co-occurrence signals and non-stop words. In addition to this strategy, we try another QTP method based on a self-attention scores matrix of query tokens' embeddings, similar to Equation~\eqref{eq:att}. We argue that, on the one hand, the token with a large value of average self-attention score represents the main semantics of the query, and therefore, would play a major role in retrieving relevant documents. On the other hand, the tokens with low self-attention scores may also convey some salient information that is different from other concrete query tokens but may be useful in retrieval. Therefore, in order to retain most of the information in the original query, we combine both the tokens with max self-attention scores and those with minimum self-attention scores.

\section{Experiment}\label{sec:experiment}

In this section, we describe the experimental settings, including the datasets, baselines, and the implementation details of training and evaluation.

\subsection{Datasets}

First, we conduct experiments on a large-scale ad-hoc retrieval dataset MS MARCO passage~\cite{bajaj2016msmarco}. This dataset consists of 8.8M passages and 800K queries, of which 500K queries are labeled with one or a few relevant passages. We train our models, along with the baseline models, on these labeled data and evaluate them in the MS MARCO passage full ranking task and TREC 2019 DL task. The former task contains 6,980 test queries and needs the model to retrieve a candidate set consisting of the top 1000 passages sorted by relevance from 8.8M passages corpus for each test query. The test set of TREC 2019 DL contains 200 queries, of which 43 have human relevance labels. For evaluation metrics, in addition to using commonly used MRR@10 for the MS MARCO passage ranking task and NDCG@10 for the TREC 2019 DL task, we also report Recall@100 for both two benchmarks. We choose Recall@100 instead of the usually used Recall@100 because, for the state-of-the-art transformer-based reranker~\cite{nogueira2019passage}, the latency of reranking 1,000 candidates is unacceptable for the online systems.

Second, we also conduct experiments on BEIR~\cite{thakur2021beir}, which is a heterogeneous zero-shot retrieval benchmark containing 18 datasets from diverse text retrieval tasks and domains. Most datasets in BEIR do not have a training set, so it is widely used for zero-shot retrieval which can evaluate models' robustness. A common pipeline is training the models on a large supervised dataset such as MS MARCO and then evaluating the models on BEIR, and we also follow this way. Following the previous work~\cite{formal2021spladev2, santhanam2022colbertv2}, we conduct our evaluation only using the 13 publicly available datasets in BEIR. For evaluation metrics of BEIR, we report NDCG@10 to evaluate the gap between the ranking of the retrieved documents and the ideal ranking of documents.

\subsection{Baselines}

We compare and analyze several typical late-interaction models, including soft-matching models and hard-matching models. Additionally, we encompass sparse retrieval models for comparison, as they can be regarded as a scalar version of hard-matching models.

\paragraph{Soft-matching late-interaction models.} ColBERT~\cite{khattab2020colbert} is a type of accurate and fast late-interaction model using a soft-matching mechanism. ColBERTv2~\cite{santhanam2022colbertv2} is an improved version of ColBERT, using an aggressive residual compression mechanism and distillation strategy to improve the index storage and retrieval performance. As these optimization techniques and the pruning methods we proposed are orthogonal, we also include ColBERTv2 as one of the baselines for experimentation.

\paragraph{Hard-matching late-interaction models.} COIL~\cite{gao2021coil} is an exact-matching based late-interaction model. COIL incorporates vector representation into the inverted index structure to accomplish exact token matching on the query and document. We also consider COIL-tok, which is COIL without [CLS] embeddings, evaluated to investigate the [CLS] embedding under various document token pruning settings.

\paragraph{Neural sparse retrieval models.} DeepImpact~\cite{mallia2021learning} is ColBERT followed by a two-layer MLP to convert it into a scalar version and perform an exact match based on an inverted index. uniCOIL~\cite{lin2021few} is a scalar version of COIL which sets the dimension of COIL's embeddings to 1. Unlike the previous two models mapping each input token to a term impact, SPLADE~\cite{formal2021splade, formal2021spladev2} is an alternative sparse retrieval model that projects dense vectors onto the dimensionality of the vocabulary and aggregates them to obtain term impacts. To ensure a comprehensive comparison, we still consider incorporating it as a baseline but only use a popular version of SPLADE, i.e., SPLADE-max, which uses a max pooling strategy.

\subsection{Implementation Details}\label{sec:implementation}

\subsubsection{Model Training}

For ColBERT\footnote{\url{https://github.com/stanford-futuredata/ColBERT}}, COIL\footnote{\url{https://github.com/luyug/COIL}}, and SPLADE\footnote{\url{https://github.com/naver/splade}}, we mainly follow the instructions provided by the official repositories and use the default settings to train our models. 

Specifically, when training ColBERT, following ~\citet{khattab2020colbert}, we use the official triples file\footnote{\url{https://microsoft.github.io/msmarco/}}. Every Query is padded or truncated to 32 tokens and passages are truncated to the first 180 tokens. The final embedding dimension of all tokens is projected to 128 which is different from the default 768 dimensions. As for COIL, passages are truncated to the first 180 tokens. The dimension of [CLS] is 768 and the dimension of the other tokens' embeddings is projected to 32. Following ~\citet{gao2021coil}, we sample 7 hard negatives from BM25's top1000 results for each query to train the model. Additionally, we use the efficient in-batch negative training method~\cite{zhan2021optimizing, huang2020embedding}.

For DeepImpact\footnote{\url{https://github.com/DI4IR/SIGIR2021}} and uniCOIL\footnote{\url{https://github.com/luyug/COIL/tree/main/uniCOIL}}, different from the original papers, we do not use a generative model to expand the original document so that we can compare the vector version model and scalar version model fairly.

\subsubsection{Data Preprocessing}

Before indexing, we first compute the document frequency of different tokens in the MS MARCO passage dataset and BEIR dataset. After getting the IDF value of each token, we list them in descending order by their IDF values (i.e., ascending order by their document frequency values) and store the sorted indices in a file. We can directly read this file when using the model to encode corpus which can avoid sorting tokens in real time so that we can save a lot of time. The data preprocessing in this work is mainly for IDF-Top-$\alpha$ pruning although we can directly integrate it into the indexing without an external file, but, that will cause much time waste.

\subsubsection{Indexing and Evaluation}

Once the training process is completed, we can utilize these models to encode each passage, while employing the proposed pruning methods. For ColBERT and COIL, we do pruning while encoding, and save the remaining embeddings with float16 data type. For ColBERT, we use Faiss to build an IVFPQ index for all embeddings, and the index is trained on a randomly selected 5\% sample of the document embeddings, while COIL is based on a special inverted index of embeddings. For all sparse model baselines, we use the Pyserini toolkit~\cite{lin2021pyserini} to build the inverted index and perform the exact match retrieval. It should be noted that due to the different model architectures of SPLADE, we employ a unique approach of pruning the output of the MLM head before performing pooling operations.

Training and evaluating are conducted on a server equipped with 2 RTX 3090 GPUs and each of them has 24GiBs of GPU memory.

\section{Results}\label{sec:results}

With the experimental results in the MS MARCO passage ranking task, TREC 2019 DL task, and BEIR dataset, we investigate the effectiveness and efficiency of different document token pruning methods. We also conduct some ablation studies to show how the effectiveness and efficiency change with parameters and settings of DTP, respectively. Besides, we also show the result of QTP on ColBERT on MS MARCO. 

\subsection{Document Token Pruning}

\subsubsection{Results on MS MARCO and TREC DL} 

\begin{table*}[t]
    \centering
    \caption{\textbf{Results of Document pruning methods} with remaining ratio $\alpha=0.5/0.75$ on MS MARCO passage ranking task and TREC 2019 Deep Learning Track. The result of $\alpha=0.5$ is on the left of "/", while the result of $\alpha=0.75$ is on the right. The bolded numbers represent the optimal results achieved by applying pruning techniques in each model. $\ast$ indicates equivalent to the unpruned settings at $p<0.05$ level using the TOST test~\cite{schuirmann1987tost} with the equivalence bound $-\Delta_L = -0.05$ and $\Delta_U = 0.05$.}
    \label{tab:baselines}
    \begin{tabular} {l|ll|ll} 
    \toprule
     \multirow{2}{*}{\textbf{Models}} 
    & \multicolumn{2}{c|}{\textbf{MS MARCO Dev}} & \multicolumn{2}{c}{\textbf{TREC DL 19}} \\  
    & \textbf{MRR@10} &  \textbf{Recall@100} & \textbf{NDCG@10} & \textbf{Recall@100} \\ \hline

ColBERT                     & .363               & .874               & .711               & .460               \\
{ColBERT + First}           & $.348^{*}$/$\textbf{.360}^{*}$ & $.839^{*}$/$.866^{*}$ & $.699\mkern7.5mu$/$\textbf{.707}\mkern7.5mu$ & $.418\mkern7.5mu$/$.444\mkern7.5mu$          \\
{ColBERT + IDF-Top}         & $.341^{*}$/$\textbf{.360}^{*}$ & $.860\mkern7.5mu$/$\textbf{.875}^{*}$ & $.672\mkern7.5mu$/$.695\mkern7.5mu$ & $.442\mkern7.5mu$/$\textbf{.457}\mkern7.5mu$ \\
{ColBERT + Attention-Top}   & $.325^{*}$/$.356^{*}$ & $.832\mkern7.5mu$/$.870^{*}$ & $.667\mkern7.5mu$/$.704\mkern7.5mu$ & $.405\mkern7.5mu$/$.453\mkern7.5mu$          \\
\hline
ColBERTv2                   & .397               & $.914$             & .749               & .562               \\
{ColBERTv2 + First}         & $.377^{*}$/$\textbf{.393}^{*}$ & $.863\mkern7.5mu$/$.901^{*}$ & $.723\mkern7.5mu$/$.740\mkern7.5mu$ & $.511\mkern7.5mu$/$.550\mkern7.5mu$          \\
{ColBERTv2 + IDF-Top}       & $.326\mkern7.5mu$/$.360^{*}$ & $.838\mkern7.5mu$/$.878^{*}$ & $.702\mkern7.5mu$/$.715\mkern7.5mu$ & $.471\mkern7.5mu$/$.519\mkern7.5mu$          \\
{ColBERTv2 + Attention-Top} & $.360^{*}$/$.390^{*}$ & $.884^{*}$/$\textbf{.912}^{*}$ & $.720\mkern7.5mu$/$\textbf{.747}\mkern7.5mu$ & $.543\mkern7.5mu$/$\textbf{.566}\mkern7.5mu$ \\
\hline
COIL-full                   & .352               & .874               & .693               & .505               \\
{COIL-full + First}         & $.338^{*}$/$.348^{*}$ & $.835^{*}$/$.862^{*}$ & $.687\mkern7.5mu$/$.692\mkern7.5mu$ & $.462\mkern7.5mu$/$.492\mkern7.5mu$          \\
{COIL-full + IDF-Top}       & $.336^{*}$/$\textbf{.350}^{*}$ & $.844^{*}$/$\textbf{.874}^{*}$ & $.651\mkern7.5mu$/$.693\mkern7.5mu$ & $.462\mkern7.5mu$/$\textbf{.509}\mkern7.5mu$ \\
{COIL-full + Attention-Top} & $.345^{*}$/$\textbf{.350}^{*}$ & $.870^{*}$/$.873^{*}$ & $.689\mkern7.5mu$/$\textbf{.694}\mkern7.5mu$ & $.504\mkern7.5mu$/$.507\mkern7.5mu$          \\
\hline
COIL-tok                    & .340               & .851               & .676               & .500               \\
{COIL-tok + First}          & $.321^{*}$/$.338^{*}$ & $.800\mkern7.5mu$/$.835^{*}$ & $.659\mkern7.5mu$/$.662\mkern7.5mu$ & $.436\mkern7.5mu$/$.482\mkern7.5mu$          \\
{COIL-tok + IDF-Top}        & $.315^{*}$/$.340^{*}$ & $.824^{*}$/$.851^{*}$ & $.639\mkern7.5mu$/$\textbf{.676}\mkern7.5mu$ & $.470\mkern7.5mu$/$\textbf{.501}\mkern7.5mu$ \\
{COIL-tok + Attention-Top}  & $.336^{*}$/$\textbf{.342}^{*}$ & $.847^{*}$/$\textbf{.853}^{*}$ & $.675\mkern7.5mu$/$\textbf{.676}\mkern7.5mu$ & $.496\mkern7.5mu$/$\textbf{.501}\mkern7.5mu$ \\
\hline
DeepImpact                  & .285               & .770               & .547               & .413               \\
{DeepImpact + First}        & $.257^{*}$/$.279^{*}$ & $.684\mkern7.5mu$/$.741^{*}$ & $.511\mkern7.5mu$/$.528\mkern7.5mu$ & $.345\mkern7.5mu$/$.388\mkern7.5mu$          \\
{DeepImpact + IDF-Top}      & $.217\mkern7.5mu$/$\textbf{.283}^{*}$ & $.642\mkern7.5mu$/$\textbf{.766}^{*}$ & $.536\mkern7.5mu$/$\textbf{.563}\mkern7.5mu$ & $.360\mkern7.5mu$/$\textbf{.430}\mkern7.5mu$ \\
{DeepImpact + Attention-Top}& $.199\mkern7.5mu$/$.232\mkern7.5mu$ & $.551\mkern7.5mu$/$.639\mkern7.5mu$ & $.449\mkern7.5mu$/$.472\mkern7.5mu$ & $.285\mkern7.5mu$/$.321\mkern7.5mu$          \\
\hline
uniCOIL                     & .319               & .809               & .636               & .479               \\
{uniCOIL + First}           & $.297^{*}$/$.314^{*}$ & $.746\mkern7.5mu$/$.790^{*}$ & $.624\mkern7.5mu$/$.630\mkern7.5mu$ & $.430\mkern7.5mu$/$.465\mkern7.5mu$          \\
{uniCOIL + IDF-Top}         & $.275\mkern7.5mu$/$.312^{*}$ & $.757\mkern7.5mu$/$.803^{*}$ & $.578\mkern7.5mu$/$.632\mkern7.5mu$ & $.460\mkern7.5mu$/$\textbf{.482}\mkern7.5mu$ \\
{uniCOIL + Attention-Top}   & $.313^{*}$/$\textbf{.319}^{*}$ & $.802^{*}$/$\textbf{.809}^{*}$ & $\textbf{.645}\mkern7.5mu$/$.638\mkern7.5mu$ & $.473\mkern7.5mu$/$.476\mkern7.5mu$          \\
\hline
SPLADE                      & .340               & .871               & .682               & .487               \\
{SPLADE + First}            & $.320^{*}$/$.336^{*}$ & $.814\mkern7.5mu$/$.852^{*}$ & $.665\mkern7.5mu$/$\textbf{.681}\mkern7.5mu$ & $.447\mkern7.5mu$/$.469\mkern7.5mu$          \\
{SPLADE + IDF-Top}          & $.203\mkern7.5mu$/$.272\mkern7.5mu$ & $.614\mkern7.5mu$/$.749\mkern7.5mu$ & $.530\mkern7.5mu$/$.611\mkern7.5mu$ & $.317\mkern7.5mu$/$.397\mkern7.5mu$          \\
{SPLADE + Attention-Top}    & $.303^{*}$/$\textbf{.338}^{*}$ & $.802\mkern7.5mu$/$\textbf{.865}^{*}$ & $.642\mkern7.5mu$/$\textbf{.681}\mkern7.5mu$ & $.445\mkern7.5mu$/$\textbf{.484}\mkern7.5mu$ \\
\bottomrule
\end{tabular}
\end{table*}

For MS MARCO, we show the effectiveness of different models after different DTP methods in Table~\ref{tab:baselines} that keep 50\% and 75\% token embeddings, and the results reveal that all late-interaction models and sparse models are compatible with the three DTP methods. In the result cells, the result of $\alpha=0.5$ is on the left of "/", and the result of $\alpha=0.75$ is on the right. This subsection mainly focuses on the effectiveness of the $\alpha=75\%$ version, and we will discuss the effectiveness of different remaining ratios in the later subsection. 

Specifically, for most DTP methods, the performance decrease of both soft-matching and hard-matching late-interaction models is minimal when remaining 75\% token embeddings. Statistical tests on the results from MS MARCO and TREC DL indicate the equivalence between the pruned approach and the unpruned setting. 

For ColBERT, the weakest DTP method for MRR@10 in the MS MARCO passage ranking task is the Attention-Top pruning method, which loses 1.7\% in MRR@10. In contrast, the other two DTP methods only have a loss of less than 1\%. IDF-Top pruning method has the best performance for Recall@100 in the MS MARCO Passage Retrieval task which can have no loss. For pruning ColBERT on the TREC DL task, the First-Top DTP method outperforms Attention-Top and IDF-Top in NDCG@10, and the IDF-Top DTP method performs best in Recall@100.
As for ColBERTv2, the IDF-Top DTP method results in a notable decline in effectiveness, while the other two methods have a little difference compared to the baseline. 

COIL is a strong hybrid model that leverages [CLS] embedding to alleviate vocabulary mismatch problems and we always remain the [CLS] embedding when we perform DTP methods. As we can see, COIL is very robust to IDF-Top DTP and Attention-Top DTP. COIL after IDF-Top DTP and Attention-Top has comparable and even better effectiveness than original COIL on both tasks which means there may be some noise in the exact matching mechanism and filtering the noise brings some gain. 
COIL-tok is a modified version of COIL that does not have [CLS] embedding. The results show that COIL-tok has a consistent decrease than COIL, which can prove that [CLS] can alleviate the vocabulary mismatch problem by introducing [CLS] semantic matching. Similar to COIL, COIL-tok after IDF-Top DTP and Attention-Top have equivalent and even slightly superior effectiveness than the original COIL-tok on both tasks, indicating that there may be some noise in the exact matching process and DTP can filter the noise then result in some gain. 

However, the sparse retrieval models exhibit significantly lower robustness to pruning, with some pruning methods resulting in a substantial performance decrease.
DeepImpact has a large loss when compared to its vector counterpart (i.e., ColBERT), indicating that vectors have stronger representation capabilities. Different from ColBERT, the Attention-Top DTP method has a significant loss compared to the unpruned setting, and the IDF-Top DTP method on DeepImpact performs best in both MS MARCO and TREC. These differences between ColBERT and DeepImpact may come from different representations (i.e., vector v.s. scalar) and different matching mechanisms (i.e., soft-matching v.s. hard-matching).
The uniCOIL model can be seen as a scalar version of COIL-tok, and the aim of evaluating uniCOIL is to compare the robustness of scalar-version models and vector-version models on DTP based on an inverted index. The results suggest that uniCOIL is weaker than COIL-tok when we employ DTP on uniCOIL. Specifically, there is a gap of 2.2\% in MRR@10 when we employ IDF-top on uniCOIl while there is no loss in Recall@100 when we employ IDF-top on COIL-tok.
As for SPLADE, due to its different architecture, it has a different behavior compared to DeepImpact and uniCOIL. The IDF-Top DTP method results in a significant loss when applied to SPLADE. We suggest that this may be attributed to the loss of information due to pruning with IDF before aggregation. In contrast, because the other two DTP methods perform better, it can be inferred that the projection of tokens positioned at the beginning (or with higher attention scores) provides more informative content.

The main results on MS MARCO show that most late-interaction models are robust to most document token pruning methods, that is to say, these models can save a lot of storage space by DTP with little performance loss. More specifically, the soft-matching late-interaction vector model (ColBERT and ColBERTv2) and the hard-matching late-interaction vector exact model (COIL and COIL-tok) are particularly resistant to DTP. However, the scalar version models (DeepImpact and uniCOIL) are more vulnerable. We believe this is because vector representation models contain some duplicate information and are hence more robust to DTP. Additionally, the Attention-Top and IDF-Top pruning methods seem to be more robust across all different models.

\subsubsection{Results on BEIR}

In addition to MS MARCO and TREC DL 2019, we also conduct experiments using these DTP methods for the above models on BEIR to evaluate the out-of-domain robustness. Table~\ref{tab:baselines_BEIR} shows the effectiveness of different models after different DTP methods. We only report the results that contain 75\% token embeddings due to the paper length.

Specifically, when we keep 75\% token embeddings, for ColBERT, the worst DTP method is the Attention-Top pruning method according to the average value of these datasets, which is 0.391 in NDCG@10 and loses 7.3\% performance compared to the original ColBERT, while other two DTP methods are both slightly better than the not pruned model, this means there are some harmful data which mislead the original model. More specifically, there are five datasets (i.e., FQ, NF, SF, TC, and TO) that pruning can improve the performance, like the effectiveness after First-Top Pruning can outperform 0.27 in NDCG@10 on Trec-Covid (i.e., TC) dataset. The First pruning method only has the four best results compared to the other two pruning methods, which is the least one among the three, while it also has a good performance in other non-best datasets, so it is more robust and gets the best average performance. The Attention-Top pruning has the six best results among these three pruning methods, however, it has the worst average performance because of its inferior results in several datasets. Compared to the First-Top and Attention-Top pruning methods, the results of the IDF-Top method placed in the middle which is comparable with the First pruning method, and the average value is better than the non-pruned ColBERT.

Different from the results on MS MARCO, we found that COIL has better average performance on tested BEIR datasets than ColBERT, which may mean COIL has more robust representations and the exact matches based on the traditional inverted index also improves the robustness so it can outperform ColBERT. However, COIL is sensitive about pruning on BEIR tasks. As we can see in the average results, for COIL-full, the best pruning method is the Attention-Top, but it still loses 11.1\% NDCG@10 compared to the original COIL. Surprisingly, COIL-tok performs more robustly when applied pruning compared to COIL. Within COIL-tok, Attention-Top is the best pruning method and only has a 0.5\% NDCG@10 loss, while the other two pruning methods have much worse effectiveness than the original COIL-tok, however, still better than COIL on the same settings. Regarding the differences between the two models, we suggest this may be attributed to the fact that for COIL, the [CLS] embedding plays a main role after pruning. However, existing research has indicated that the zero-shot performance of solely using the [CLS] embedding from fine-tuned BERT is subpar~\cite{thakur2021beir}. Conversely, COIL-tok relies entirely on the exact match of token embeddings and this approach may potentially yield greater robustness in a zero-shot setting. 

As for the sparse models, the best pruning method is IDF-Top for both DeepImpact and SPLADE and is Attention-Top for uniCOIL. DeepImpact and uniCOIL have a large loss compared to ColBERT and COIL, indicating that vectors have stronger representation capabilities no matter whether in the in-domain dataset or out-domain dataset. However, SPLADE shows a comparable performance, which may derive from the different architecture.

\begin{table*}[t]
\centering
\caption{\textbf{NDCG@10 results of Document pruning methods} with remaining ratio $\alpha=0.75$ on BEIR dataset. (AR: ArguAna. CF: Climate-Fever. DB: DBPedia. FE: Fever. FQ: FiQA-2018. HQ: HotpotQA. NF: NFCorpus. NQ: Natural Questions. QU: Quora. SD: SCIDOCS. SF: Scifact. TC: TREC-COVID. TO: Touché-2020.)}
\label{tab:baselines_BEIR}
\footnotesize
\begin{tabular}{l|c|c|c|c|c|c|c|c|c|c|c|c|c|c}
\toprule
                       & \textbf{AR} & \textbf{CF} & \underline{\textbf{DB}} & \textbf{FE} & \underline{\textbf{FQ}} & \underline{\textbf{HQ}} & \underline{\textbf{NF}} & \underline{\textbf{NQ}} & \textbf{QU} & \textbf{SD} & \textbf{SF} & \underline{\textbf{TC}} & \underline{\textbf{TO}} & \textbf{Avg.} \\
    \hline  
    \textbf{ColBERT}   & .398 & .148 & .377 & .747 & .299 & .570 & .291 & .516 & .841 & .145 & .604 & .600 & .205 & .422 \\
    \textbf{First}     & .387 & .138 & .367 & .717 & .289 & .527 & .290 & \textbf{.499} & .788 & \textbf{.141} & .580 & \textbf{.627} & \textbf{.211} & \textbf{.428} \\
    \textbf{IDF-Top}   & .390 & .138 & .371 & .719 & \textbf{.299} & .534 & \textbf{.293} & .399 & \textbf{.834} & \textbf{.141} & \textbf{.612} & .600 & \textbf{.211} & .426 \\
    \textbf{Att-Top}   & \textbf{.391} & \textbf{.146} & \textbf{.373} & \textbf{.740} & .192 & \textbf{.561} & .190 & .385 & .732 & .098 & .497 & .443 & .203 & .391 \\
    \hline
    \textbf{ColBERTv2} & .463 & .176 & .446 & .785 & .356 & .667 & .338 & .562 & .852 & .154 & .693 & .738 & .263 & .499 \\
    \textbf{First}     & .444 & .155 & .440 & .730 & .324 & .628 & .324 & .532 & .787 & .152 & .616 & \textbf{.721} & .261 & .470  \\
    \textbf{IDF-Top}   & \textbf{.449} & \textbf{.170} &\textbf{ .451} & \textbf{.753} & \textbf{.340} & \textbf{.673} & \textbf{.326} & \textbf{.550} & \textbf{.853} & \textbf{.155} & \textbf{.648} & .717 & \textbf{.262} & \textbf{.488} \\
    \textbf{Att-Top}   & .445 & .167 & .444 & .739 & .336 & .658 & \textbf{.326} & .547 & .844 & \textbf{.155} & .640 & .700 & \textbf{.262} & .482 \\
    \hline
    \textbf{COIL}      & .372 & .149 & .393 & .751 & .268 & .593 & .307 & .598 & .827 & .149 & .555 & .631 & .240 & .449 \\
    \textbf{First}     & .327 & \textbf{.079} & .263 & .683 & .219 & .536 & .306 & .451 & .557 & \textbf{.138} & .518 & .504 & .225 & .370 \\
    \textbf{IDF-Top}   & \textbf{.342} & .072 & .246 & .605 & \textbf{.234} & .473 & \textbf{.319} & .444 & .498 & .135 & .542 & .490 & .219 & .355 \\
    \textbf{Att-Top}   & .336 & \textbf{.079} & \textbf{.270} & \textbf{.753} & .232 & \textbf{.557} & .311 & \textbf{.471} & \textbf{.754} & .135 & \textbf{.545} & \textbf{.517} & \textbf{.233} & \textbf{.399} \\
    \hline
    \textbf{COIL-tok}  & .351 & .124 & .370 & .725 & .245 & .568 & .306 & .468 & .796 & .144 & .547 & .582 & .244 & .421 \\
    \textbf{First}     & .333 & .124 & .356 & .709 & .234 & .539 & .302 & .444 & .621 & .142 & .533 & \textbf{.580} & .235 & .396 \\
    \textbf{IDF-Top}   & \textbf{.350} & .109 & .327 & .649 & \textbf{.247} & .477 & \textbf{.307} & .434 & .555 & .143 & \textbf{.550} & .568 & \textbf{.260} & .383 \\
    \textbf{Att-Top}   & \textbf{.350} & \textbf{.125} & \textbf{.368} & \textbf{.724} & .241 & \textbf{.562} & .305 & \textbf{.465} & \textbf{.791} & \textbf{.144} & .546 & .575 & .256 & \textbf{.419} \\
    \hline
    \textbf{DeepImpact}  & .375 & .043 & .167 & .027 & .108 & .092 & .232 & .329 & .248 & .090 & .390 & .336 & .195 & .202 \\
    \textbf{First}     & \textbf{.384} & .040 & .164 & \textbf{.026} & .099 & .082 & .227 & .311 & .160 & \textbf{.091} & .364 & .322 & \textbf{.190} & .189 \\
    \textbf{IDF-Top}   & .372 & \textbf{.060} & \textbf{.165} & .021 & \textbf{.121} & .075 & \textbf{.287} & .312 & \textbf{.286} & .080 & .362 & .329 & .152 & \textbf{.202} \\
    \textbf{Att-Top}   & \textbf{.384} & .039 & \textbf{.165} & .025 & .102 & \textbf{.083} & .232 & \textbf{.315} & .231 & .089 & \textbf{.382} & \textbf{.332} & .188 & .197 \\
    \hline
    \textbf{uniCOIL}   & .330 & .107 & .340 & .727 & .226 & .550 & .302 & .413 & .650 & .136 & .527 & .561 & .241 & .393 \\
    \textbf{First}     & .315 & .107 & .330 & .709 & .213 & .525 & .298 & .391 & .435 & .134 & .506 & .564 & .239 & .367 \\
    \textbf{IDF-Top}   & \textbf{.334} & .101 & .302 & .643 & .226 & .457 & .302 & .379 & .406 & .136 & .527 & .549 & \textbf{.254} & .355 \\
    \textbf{Att-Top}   & .329 & \textbf{.109} & \textbf{.340} & \textbf{.727} & \textbf{.226} & \textbf{.547} & \textbf{.302} & \textbf{.413} & \textbf{.646} & \textbf{.137} & \textbf{.529} & \textbf{.566} & .241 & \textbf{.393} \\
    \hline
    \textbf{SPLADE}    & .439 & .119 & .366 & .730 & .287 & .636 & .313 & .469 & .835 & .145 & .628 & .673 & .256 & .456 \\
    \textbf{First}     & .454 & .154 & .366 & .678 & .235 & .610 & .303 & .439 & .693 & .144 & .555 & .632 & .184 & .419 \\
    \textbf{IDF-Top}   & \textbf{.479} & .159 & \textbf{.368} & .685 & \textbf{.250} & \textbf{.642} & \textbf{.308} & .451 & \textbf{.829} & \textbf{.148} & \textbf{.581} & .631 & .199 & \textbf{.441} \\
    \textbf{Att-Top}   & .477 & \textbf{.161} & \textbf{.368} & \textbf{.694} & .248 & .639 & .306 & \textbf{.454} & .635 & .146 & .580 & \textbf{.634} & \textbf{.220} & .428 \\
\bottomrule
\end{tabular}
\end{table*}

Additionally, we analyze the pruning methods across the different types of datasets. Following ~\citet{santhanam2022colbertv2}, we classify the 13 datasets into two types: search tasks and semantic relatedness tasks. The first type of dataset includes DB, FQ, HQ, NF, NQ, and TC, and the second type includes AR, CF, FE, QU, SD, and SF. The datasets of search tasks are underlined in Table~\ref{tab:baselines_BEIR}.

We count the number of datasets in which different pruning methods achieved the best performance on the two types of datasets, and the results are shown in Table~\ref{tab:beir_win}. We can see that the First pruning is the weakest, which is consistent with the results on MS MARCO. Comparing the other two pruning methods, we find that for soft-matching models (i.e., ColBERT and ColBERTv2), the IDF-Top pruning seems to be better, while Attention-Top pruning achieves better performance for other models based on exact match. This result is surprising because IDF value is first designed for the inverted index which is also based on exact match. We think this may attributed to the difference in the distribution between IDF values and learned impact or embeddings. However, there is no obvious conclusion can be drawn about which pruning method is preferred for different types of datasets.

\begin{table}[]
\caption{The number of datasets in which different pruning methods achieved the best
performance on the two types of datasets. \textbf{S} indicates search tasks including 6 datasets and \textbf{SR} indicates semantic relatedness tasks including 7 datasets. Note that since there are cases where the results of both methods are the same, the sum of each column may exceed the number of datasets.}
\label{tab:beir_win}
\begin{tabular}{@{}l|cc|cc|cc|cc|cc|cc|cc@{}}
\toprule
                     \multirow{2}{*}{\textbf{DTP Method}} 
                   & \multicolumn{2}{c|}{\rotatebox{45}{\textbf{ColBERT}}} 
                   & \multicolumn{2}{c|}{\rotatebox{45}{\textbf{ColBERTv2}}} 
                   & \multicolumn{2}{c|}{\rotatebox{45}{\textbf{COIL}}} 
                   & \multicolumn{2}{c|}{\rotatebox{45}{\textbf{COIL-tok}}} 
                   & \multicolumn{2}{c|}{\rotatebox{45}{\textbf{DeepImpact}}} 
                   & \multicolumn{2}{c|}{\rotatebox{45}{\textbf{uniCOIL}}} 
                   & \multicolumn{2}{c}{\rotatebox{45}{\textbf{SPLADE}}} \\ 
                   & S & SR & S & SR & S & SR & S & SR & S & SR & S & SR & S & SR \\
\midrule
\textbf{First}     & 3 & 1 & 1 & 0 & 0 & 2 & 1 & 0 & 1 & 3 & 0 & 0 & 0 & 0             \\
\textbf{IDF}       & 3 & 3 & 5 & 7 & 2 & 1 & 3 & 2 & 3 & 2 & 1 & 1 & 4 & 4             \\
\textbf{Attention} & 2 & 3 & 2 & 1 & 5 & 4 & 3 & 5 & 4 & 2 & 6 & 5 & 4 & 2             \\ \bottomrule
\end{tabular}
\end{table}

The main results on BEIR show some different results compared to MS MARCO. Specifically, COIL-full is the best model on the BEIR task, and it reveals that COIL-full has more robust representations and the exact matches based on the traditional inverted index also improve the robustness. Besides, the First-Top DTP method on ColBERT gives the best performance which means we can use the First pruning method to filter some harmful information. The best DTP method for all three COIL-related models is Attention-Top and it may mean this method can extract better representations more effectively.

\subsection{Query Token Pruning}

We show the results of query token pruning in Table \ref{tab:QTP}. The motivation for query token pruning is to speed up retrieval without significantly affecting performance. Therefore, we report the performance and latency of the original ColBERT model without query token pruning and with query token pruning. In addition to IDF-Top QTP used in ~\cite{tonellotto2021query}, we also evaluate the attention-based query token pruning method which is only based on the information of the given query. 

For the original ColBERT, each query retrieves 5,282 documents on average, while after IDF QTP, the average number of documents retrieved by each query is reduced to 1,972. This reduces retrieval time from 125ms to 86ms, and the speed is increased by $1.45\times$ times with no performance loss on MRR@10. After Attention QTP, the average number of documents retrieved by each query is reduced to 2,731. This leads to a reduction of retrieval time from 125ms to 98ms, and the speed is increased by $1.27\times$ times and also with no performance loss on MRR@10. 

The results can be comprehended with relative ease: ColBERT employs a two-step retrieval process. Initially, it retrieves an approximate set of candidates, followed by utilizing the sum-of-max operation to accurately compute the relevance score and rerank the candidates accordingly. Our QTP methods only prune the query tokens during the initial stage, while employing all token embeddings to compute the final score. Consequently, once the relevant document is retrieved as a candidate, its final score will be identical to that of the no QTP settings. Thus, these findings indicate that when ColBERT performs well (at the very least, for queries with MRR@10 > 0), the addition of QTP does not lead to a performance decrease on MRR. However, it is worth noting that QTP may result in a decrease in Recall@100 as the average number of retrieved documents is substantially smaller than the no QTP settings.

Furthermore, we have conducted additional experiments to confirm whether the application of QTP exhibits similar behavior when on the index constructed using the DTP method. We choose the pruned ColBERT index established using the First DTP method because it is one of the best DTP methods, while the remaining ratio $\alpha$ is set to 50\%, denoted as \textbf{ColBERT-First-50\%}. As shown in Table~\ref{tab:QTP}, the results of QTP on the pruned index are similar to those on the full index. Specifically, IDF QTP and Attention QTP both lead to a reduction in retrieval time and increase the retrieval speed by $1.31\times$ times and $1.22\times$ times, respectively, with little performance loss.

In summary, DTP actually not only reduces storage overhead but also reduces latency because there are fewer embeddings that need the model to traverse. Additionally, QTP can further improve the efficiency of the models with little loss of retrieval effectiveness.

\begin{table}[t]
    \centering
    \caption{Results of query tokens pruning on MS MARCO Dev. A.R.D. means average retrieved documents per query. Latency is the average of all query latency. $\ast$ indicates equivalent to the unpruned settings at $p<0.05$ level using the TOST test~\cite{schuirmann1987tost} with the equivalence bound $-\Delta_L = -0.05$ and $\Delta_U = 0.05$.}
    \begin{tabular}{l|cc|cc}
    \toprule
    \textbf{Models}
    & \textbf{MRR@10} & \textbf{Recall@100} &\textbf{A.R.D.} & \textbf{Latency (ms)}
    \\ \hline 
    \textbf{ColBERT}                   & $0.363\mkern7.5mu$ & $0.874\mkern7.5mu$ & 5282 & 125 \\
    {+ IDF~\cite{tonellotto2021query}} & $0.363^{*}$ & $0.860^{*}$ & 1972 & 86 \\
    {+ Attention}                      & $0.363^{*}$ & $0.868^{*}$ & 2731 & 98 \\
    \hline
    \textbf{ColBERT-First-50\%}        & $0.348\mkern7.5mu$ & $0.838\mkern7.5mu$ & 5856 & 94 \\
    {+ IDF~\cite{tonellotto2021query}} & $0.348^{*}$ & $0.829^{*}$ & 2210 & 72 \\    
    {+ Attention}                      & $0.348^{*}$ & $0.836^{*}$ & 3033 & 77 \\
    \bottomrule
    \end{tabular}
    \label{tab:QTP}
\end{table}

\subsection{Ablation and Analysis}

\subsubsection{Analysis of Remaining Ratio for Document Token Pruning}

\begin{figure*}[t]
\centering

\subfigure[ColBERT]
{\includegraphics[width=0.32\textwidth]{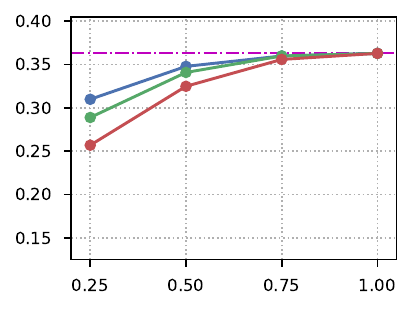}\label{subfig:ColBERT_MARCO}}
\hspace{10pt}
\subfigure[ColBERTv2]
{\includegraphics[width=0.32\textwidth]{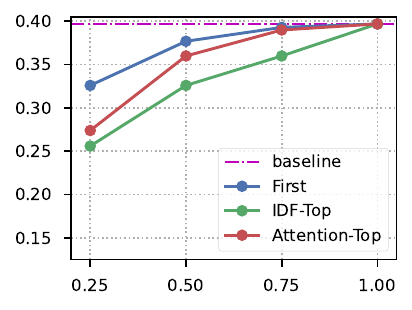}\label{subfig:ColBERTv2_MARCO}}

\subfigure[COIL]
{\includegraphics[width=0.32\textwidth]{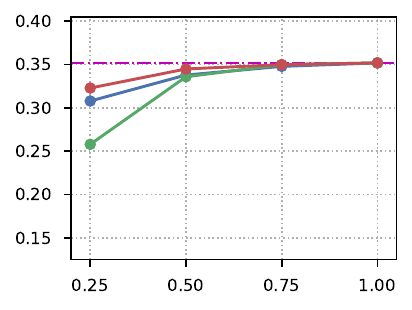}\label{subfig:COIL_MARCO}}
\hspace{10pt}
\subfigure[COIL-tok]
{\includegraphics[width=0.32\textwidth]{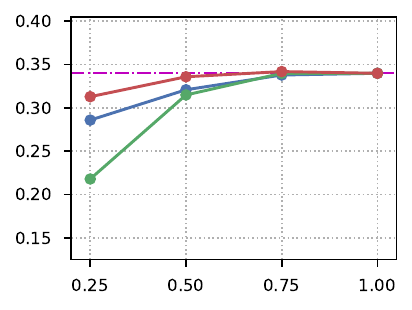}\label{subfig:COIL-tok_MARCO}}

\subfigure[DeepImpact]
{\includegraphics[width=0.32\textwidth]{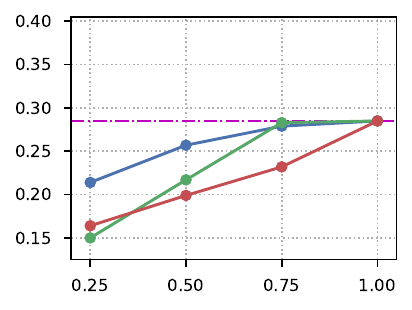}\label{subfig:DeepImapct_MARCO}}
\subfigure[uniCOIL]
{\includegraphics[width=0.32\textwidth]{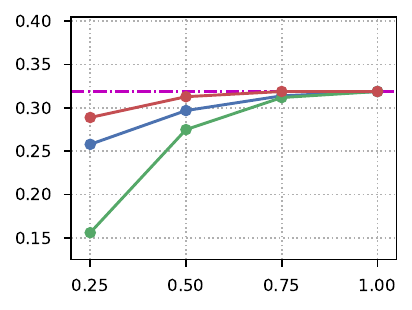}\label{subfig:uniCOIL_MARCO}}
\subfigure[SPLADE]
{\includegraphics[width=0.32\textwidth]{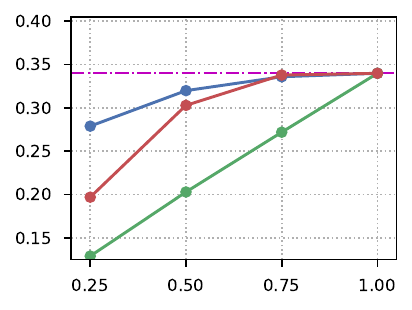}\label{subfig:SPLADE_MARCO}}

\caption{Effectiveness of different DTP methods of different models on MSMARCO Dev. The dashed line denotes the performance of models without DTP. The x-axis is the remaining ratio $\alpha$ and the y-axis of each sub-figure is the MRR@10 metric.}
\label{fig:dtp_abl_marco}
\end{figure*}

\paragraph{Analysis on MS MARCO} 

We report the results of the ablation study in Figure~\ref{fig:dtp_abl_marco}. We did a series of ablation experiments on the MS MARCO to explore the impact of the remaining ratio of tokens on the final performance of different late-interaction models. We prune on different remaining ratios and verify the effectiveness of different models. In general, the result shows that the performance of models is positively correlated with the number of representations, while the correlation is not linear which means we can further make an effectiveness-storage trade-off, for example, we can use 50\% or 75\% tokens to represent the document with only a little performance loss. However, if we use only 25\% tokens, the effectiveness of all models will have a relatively obvious decrease.

Specifically, for the results of three DTP methods on soft-matching late-interaction models (i.e., ColBERT and ColBERTv2) shown in Figure~\ref{subfig:ColBERT_MARCO} and Figure~\ref{subfig:ColBERTv2_MARCO}, the First pruning method is the best for different remaining ratios. In contrast, the IDF-Top pruning and Attention-Top pruning perform differently on the two models. It's worth noting that the IDF-Top pruning method leads to a significant performance loss even if we keep 75\% embeddings. 

As for the results of three DTP methods on hard-matching late-interaction models (i.e., COIL and COIL-tok) shown in Figure~\ref{subfig:COIL_MARCO} and Figure~\ref{subfig:COIL-tok_MARCO}, except for IDF-Top pruning, the other two pruning methods show a better robustness, even thought for remaining only 25\% embeddings. Concretely, the Attention-Top method outperforms the First method, while the First method outperforms the IDF-Top method. We also notice that the curve of COIL-tok decreases faster as the remaining ratio decreases, we posit that it's because of the lack of [CLS] embedding.

Sparse retrieval models (i.e., DeepImpact, uniCOIL, and SPLADE) hold a similar conclusion to hard-matching models, that is, the First pruning method and the Attention-Top pruning method still have good robustness while we remain more tokens (i.e., $\alpha = 0.5/0.75$), and the only exception is the Attention-Top on DeepImpact. We guess this is because DeepImpact was not trained enough. However, the curve of IDF-Top pruning on all three models decreases fast as the remaining ratio decreases to 25\%, showing that it's not enough to capture all information of a passage in an inverted index using only a small proportion of words that have the highest IDF values.

In summary, for various models, the First pruning method demonstrates relatively robust performance. In other words, as the proportion of retained tokens gradually decreases, the decline in retrieval performance is not as rapid, and it is possible to achieve performance comparable to that of unpruned settings when the remaining rate is 75\%. The other two pruning methods have different performances on different models, but the Attention-Top pruning method shows a more promising robustness.

\begin{figure*}[t]
\centering
\subfigure[COIL-tok]{\includegraphics[width=0.45\textwidth]{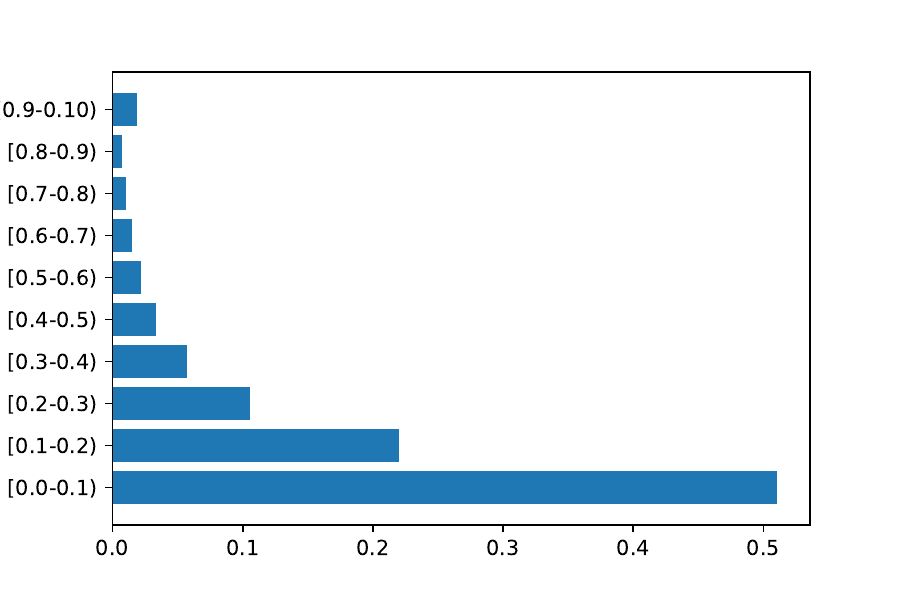}\label{subfig:Att_COIL}}
\subfigure[ColBERT]{\includegraphics[width=0.45\textwidth]{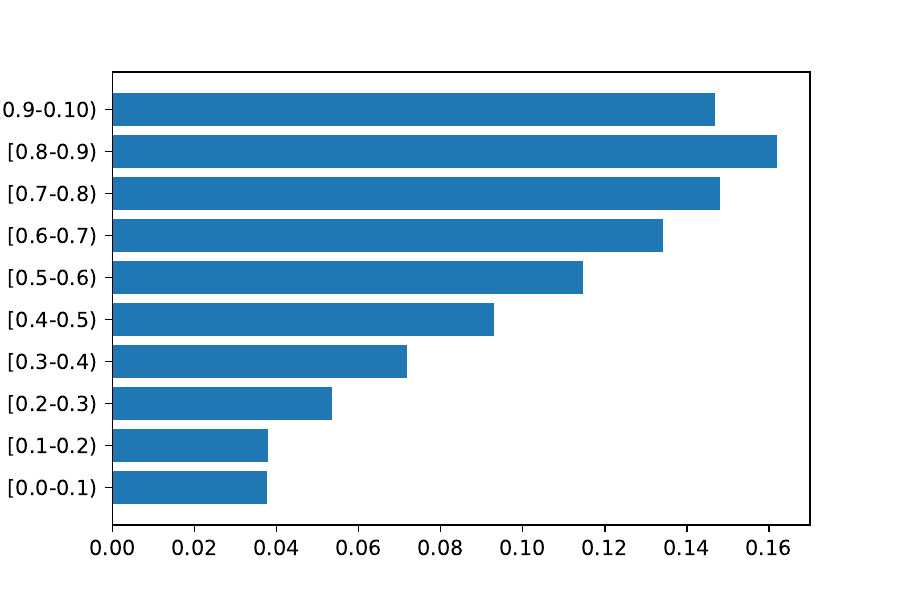}\label{subfig:Att_ColBERT}}
\caption{The self-attention distribution of COIL-tok and ColBERT. We use softmax to scale all values to [0,1] and divide them into ten bins.}
\label{fig:inner_att}
\end{figure*}

\begin{figure*}[t]
\centering
\includegraphics[width=\textwidth]{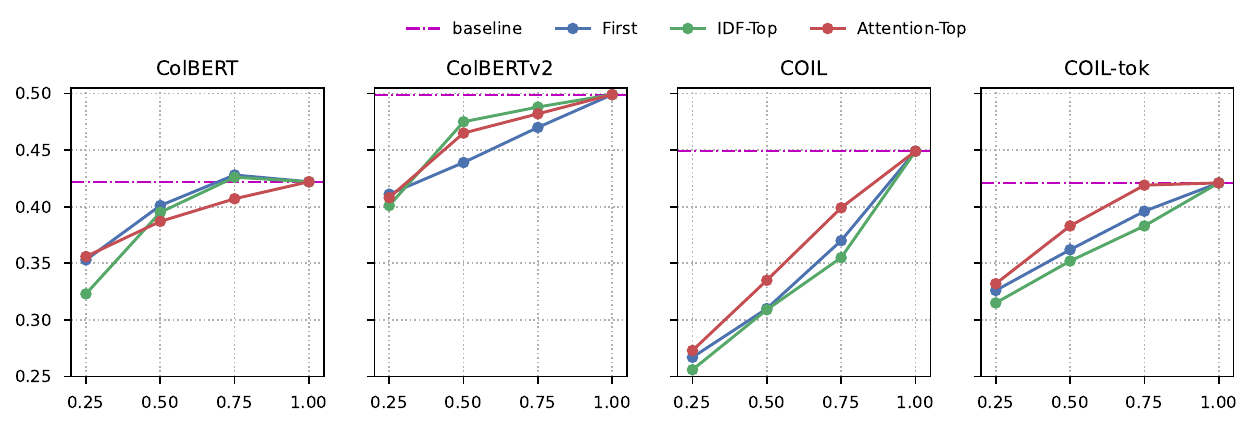}
\caption{Average effectiveness of different DTP methods on different models on BEIR. The dashed line denotes the performance of late-interaction models without DTP. The x-axis is the remaining ratio $\alpha$ and the y-axis of each sub-figure is the average NDCG@10 metric on BEIR.}
\label{fig:dtp_abl_beir}
\end{figure*}

Additionally, we also do a comparison across all models, and we found that ColBERT only outperforms DeepImpact and SPLADE and is inferior to the other three models on MS MARCO Dev when we only keep 25\% token embeddings using Attention-Top-$\alpha$ pruning which is shown by Figure~\ref{fig:dtp_abl_marco}. And if we keep more embeddings such as 75\%, the effectiveness of ColBERT will reach the top place. The reason may be the distribution of ColBERT's representation is more compact than other models, so the Attention-Top-25\% is somewhat redundant while embeddings of other models may be more scattered, so the Attention-Top-25\% embeddings of them are more informative. To confirm this, we visualize the distribution of COIL-tok and ColBERT self-attention score which is shown in Figure~\ref{fig:inner_att}. The figure tells us that the distribution of ColBERT's representation is more compact than COIL-tok's. Specifically, ColBERT's self-attention score is largely concentrated in areas greater than 0.5 whereas most COIL-tok's self-attention score is less than 0.2 which means it is scattered in the representation space. As a result, the attention-top-25\% embeddings of COIL-tok can get comparable performance with COIL-tok without pruning, however, the attention-top-25\% embeddings of ColBERT have a significant loss in most metrics.

\paragraph{Analysis on BEIR task} 

In addition to experiments on large-scale supervised datasets MS MARCO, we also conduct experiments on BEIR, and the results are shown in Figure~\ref{fig:dtp_abl_beir}. From the results, we can find that the Attention-Top pruning method always performs well across the late-interaction models when we remain 75\% embeddings, especially for COIL and COIL-tok on which it outperforms the other two methods. The First pruning and IDF-Top pruning perform differently on soft-matching models and hard-matching models, specifically, First pruning is always better than IDF-Top pruning on COIL and COIL-tok, while each has its own winner or loser on the ColBERT and ColBERTv2. However, if we only remain 25\% embeddings, all three pruning methods will increase fast and lead to a significant NDCG@10 decrease for all models.

\begin{figure*}[t]
\centering
\subfigure[ColBERT]{\includegraphics[width=0.35\textwidth]{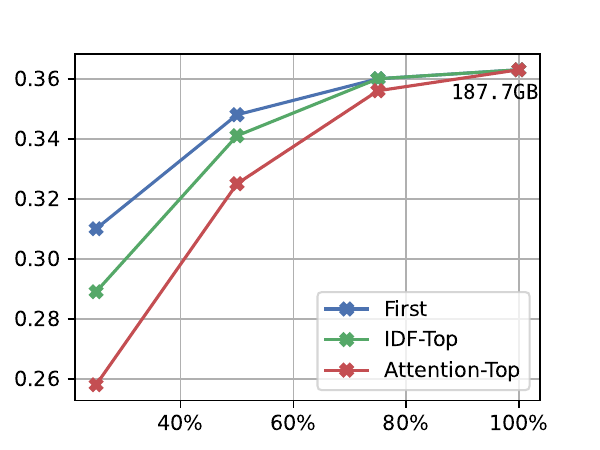}\label{subfig:tradeoff_ColBERT}}
\hspace{1cm}
\subfigure[ColBERTv2]{\includegraphics[width=0.35\textwidth]{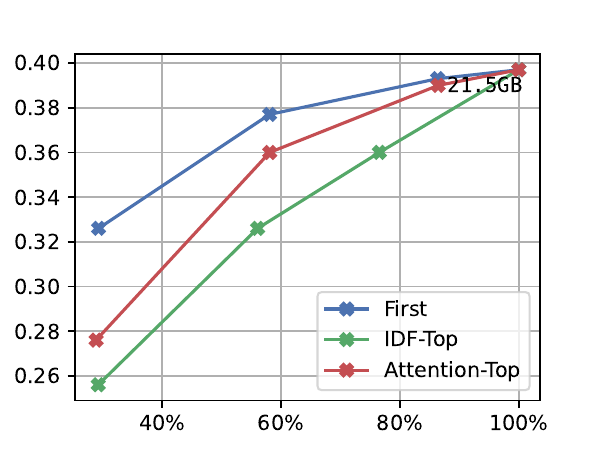}\label{subfig:tradeoff_ColBERTv2}}

\subfigure[COIL]{\includegraphics[width=0.35\textwidth]{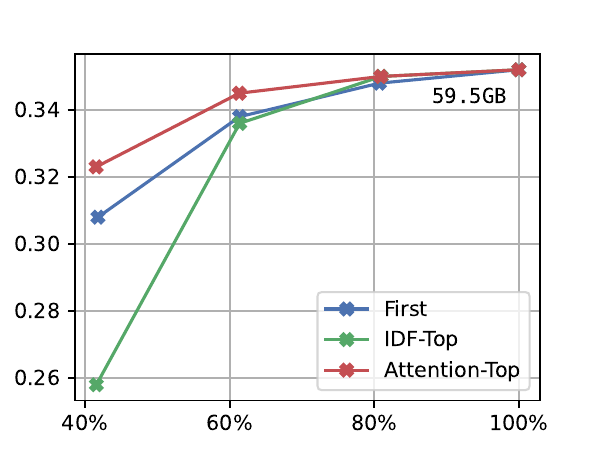}\label{subfig:tradeoff_COIL}}
\hspace{1cm}
\subfigure[COIL-tok]{\includegraphics[width=0.35\textwidth]{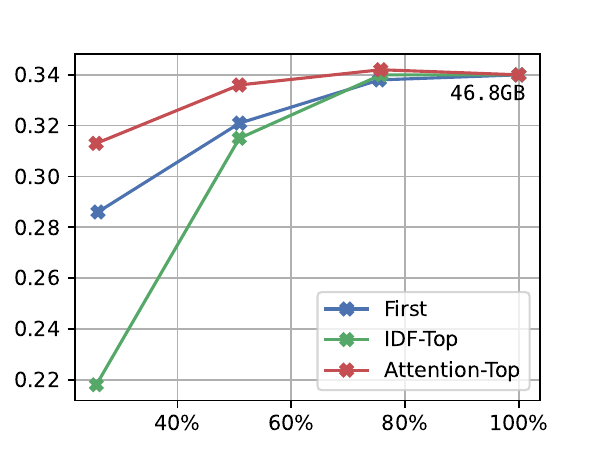}\label{subfig:tradeoff_COIL-tok}}
\caption{Index size vs. MRR@10 on MSMARCO Dev for the three pruning methods. The number on the figure is the size of the full index. The x-axis denotes the percentage of the pruned index size in relation to the full index size, while the y-axis represents the MRR@10.}
\label{fig:tradeoff}
\end{figure*}

\paragraph{Analysis of the Trade-off between Effectiveness and Efficiency} Figure~\ref{fig:tradeoff} shows a trade-off between the effectiveness (i.e., the retrieval performance on MS MARCO) and the storage efficiency (i.e., the index size). As the figure shows, we can achieve a good trade-off by remaining 75\% of token embeddings, reducing a lot of storage overhead yet with little performance loss.

\subsubsection{Analysis of Query Token Pruning}

For the method of pruning query tokens based on attention, we investigate various configurations. Specifically, we use three ways of embedding query tokens to retrieve documents: tokens with the minimum 3 attention scores, tokens with the maximum 3 attention scores, and a combination of the first two. The experiment results are shown in Table~\ref{tab:QTP_abl}. 

For the results, if we only use the 3 tokens with maximum attention scores, there will be a large performance loss compared to the original ones. Observing the much smaller average number of retrieved documents, we believe that these tokens do not capture the diversity distribution of the original query representations. Moreover, we also use the 3 token embeddings with minimum attention scores to retrieve the document candidates. Surprisingly, as the result shows, we can use these token embeddings to retrieve most of the useful documents that heavily influence the final ranking performance. We will qualitatively analyze this phenomenon later.

Moreover, we combine the minimum and maximum attention query tokens to retrieve more documents, the results show that it takes only a little improvement but more latency.

\begin{table}[t]
    \centering
    \caption{Results of query tokens pruning on MS MARCO Dev. A.R.D. means average retrieved documents per query. Latency is the average of all query latency. Min(Max)A-$k$ means using $k$ tokens of minimum (maximum) attention scores to retrieve candidate documents. }
    \begin{tabular}{l|cc|cc}
    \toprule
    \textbf{Model} & \textbf{MRR@10} & \textbf{Recall@100} & \textbf{A.R.D.} & \textbf{Latency (ms)}
    \\ \hline
    ColBERT    & 0.363 & 0.874 & 5282 & 125 \\
    {+ MinA-3} & 0.363 & 0.856 & 1952 & 99 \\
    {+ MaxA-3} & 0.340 & 0.631 & 1527 & 85 \\
    {+ MinA-3+MaxA-3} & 0.363 & 0.868 & 2731 & 98 \\
    \hline
    ColBERT-First-50\% & 0.348 & 0.836 & 5856 & 94 \\
    {+ MinA-3} & 0.347 & 0.822 & 2191 & 72 \\
    {+ MaxA-3} & 0.326 & 0.608 & 1664 & 69 \\
    {+ MinA-3+MaxA-3} & 0.348 & 0.836 & 3033 & 78 \\
    
    \bottomrule
    \end{tabular}
    \label{tab:QTP_abl}
\end{table}

\subsubsection{Qualitative Analysis and Case Study}

In order to gain a deeper comprehension of the pruning process, we qualitatively analyze the document tokens stored by ColBERT and COIL after pruning. Specifically, we only analyze the Attention-Top pruning method, as the results of the other two pruning methods are easy to obtain. In Table~\ref{tab:case_study}, we select one typical example to demonstrate how the attention scores of tokens in documents are distributed.

One of the important findings is that COIL gives higher attention scores to tokens that appear in both the query and document (i.e., ``\textit{north}'' and ``\textit{salem}'' in this example), but ColBERT does the opposite. This may explain the phenomenon that COIL performs well for Attention-Top pruning even though only remained 25\% tokens but ColBERT performs poorly, as our analysis shows that the contribution of co-concurrence is important for ColBERT (reference Table~\ref{tab:contribution}).

Another observation is that both COIL and ColBERT give some stop words high attention scores. However, our previous experimental results (reference Table~\ref{tab:contribution} and the part of IDF-Top pruning on Table~\ref{tab:baselines}) indicate that stop-words are not essential in the matching mechanism of late-interaction models. This situation may be one of the factors limiting the robustness of Attention-Top pruning.

Additionally, we compute the attention scores of the query tokens for ColBERT. We find that this distribution is similar to the attention score of tokens in the document, that is, the special tokens and stop words hold the highest attention score but the tokens with co-concurrence signals hold the lowest attention score. This finding may potentially explain the better retrieval performance observed in QTP for tokens with lower attention scores.

\begin{table}[t]
  \caption{Visualization of document token attention scores. Red shades reflect the tokens' relative ranking of attention scores, the darker the color, the higher the ranking. Query tokens are bold. The query tokens marked blue are the three tokens with the highest attention scores calculated using Colbert, and the ones marked orange are the three lowest.}
  \label{tab:case_study}
  \begin{tabular}{p{1.8cm}p{11cm}}
  \toprule
  & \colorbox{red!60}{top 25\%} \colorbox{red!40}{top 50\%} \colorbox{red!20}{ top 75\%} \colorbox{red!0}{top 100\%} \\ \midrule
  \textbf{Query} & \textbf{\textcolor{blue}{[CLS]} \textcolor{blue}{[Q]} \textcolor{orange}{average} \textcolor{orange}{income} \textcolor{blue}{for} \textcolor{orange}{north} salem new york} \\ \midrule
  
  \textbf{Document} (ColBERT)  & \colorbox{red!20}{teacher} \colorbox{red!0}{\textbf{north}} \colorbox{red!0}{\textbf{salem}} \colorbox{red!60}{,} \colorbox{red!0}{ny} \colorbox{red!40}{salary} \colorbox{red!0}{.} \colorbox{red!20}{teacher} \colorbox{red!0}{\textbf{north}} \colorbox{red!0}{\textbf{salem}} \colorbox{red!60}{,} \colorbox{red!0}{ny} \colorbox{red!40}{\textbf{average}} \colorbox{red!40}{salary} \colorbox{red!60}{is} \colorbox{red!40}{\$} \colorbox{red!60}{74} \colorbox{red!20}{,} \colorbox{red!40}{97} \colorbox{red!60}{\#\#3} \colorbox{red!20}{,} \colorbox{red!40}{median} \colorbox{red!40}{salary} \colorbox{red!60}{is} \colorbox{red!40}{\$} \colorbox{red!60}{80} \colorbox{red!0}{,} \colorbox{red!60}{000} \colorbox{red!60}{with} \colorbox{red!60}{a} \colorbox{red!40}{salary} \colorbox{red!40}{range} \colorbox{red!60}{from} \colorbox{red!40}{\$} \colorbox{red!60}{49} \colorbox{red!0}{,} \colorbox{red!40}{92} \colorbox{red!40}{\#\#0} \colorbox{red!60}{to} \colorbox{red!40}{\$} \colorbox{red!60}{86} \colorbox{red!0}{,} \colorbox{red!40}{40} \colorbox{red!40}{\#\#6} \colorbox{red!20}{.} \colorbox{red!20}{teacher} \colorbox{red!0}{\textbf{north}} \colorbox{red!0}{\textbf{salem}} \colorbox{red!60}{,} \colorbox{red!0}{ny} \colorbox{red!40}{salaries} \colorbox{red!20}{are} \colorbox{red!0}{collected} \colorbox{red!20}{from} \colorbox{red!20}{government} \colorbox{red!0}{agencies} \colorbox{red!20}{and} \colorbox{red!0}{companies} \colorbox{red!60}{.} \colorbox{red!60}{each} \colorbox{red!20}{salary} \colorbox{red!20}{is} \colorbox{red!0}{associated} \colorbox{red!20}{with} \colorbox{red!20}{a} \colorbox{red!20}{real} \colorbox{red!20}{job} \colorbox{red!20}{position} \colorbox{red!0}{.} \\ \midrule
  
  \textbf{Document} (COIL)  & \colorbox{red!20}{teacher} \colorbox{red!60}{\textbf{north}} \colorbox{red!60}{\textbf{salem}} \colorbox{red!60}{,} \colorbox{red!0}{ny} \colorbox{red!60}{salary} \colorbox{red!60}{.} \colorbox{red!0}{teacher} \colorbox{red!60}{\textbf{north}} \colorbox{red!60}{\textbf{salem}} \colorbox{red!60}{,} \colorbox{red!0}{ny} \colorbox{red!60}{\textbf{average}} \colorbox{red!40}{salary} \colorbox{red!60}{is} \colorbox{red!20}{\$} \colorbox{red!40}{74} \colorbox{red!60}{,} \colorbox{red!20}{97} \colorbox{red!40}{\#\#3} \colorbox{red!40}{,} \colorbox{red!0}{median} \colorbox{red!40}{salary} \colorbox{red!60}{is} \colorbox{red!0}{\$} \colorbox{red!40}{80} \colorbox{red!40}{,} \colorbox{red!0}{000} \colorbox{red!20}{with} \colorbox{red!20}{a} \colorbox{red!0}{salary} \colorbox{red!60}{range} \colorbox{red!20}{from} \colorbox{red!20}{\$} \colorbox{red!40}{49} \colorbox{red!40}{,} \colorbox{red!20}{92} \colorbox{red!20}{\#\#0} \colorbox{red!20}{to} \colorbox{red!20}{\$} \colorbox{red!40}{86} \colorbox{red!40}{,} \colorbox{red!20}{40} \colorbox{red!40}{\#\#6} \colorbox{red!40}{.} \colorbox{red!0}{teacher} \colorbox{red!60}{\textbf{north}} \colorbox{red!60}{\textbf{salem}} \colorbox{red!60}{,} \colorbox{red!0}{ny} \colorbox{red!60}{salaries} \colorbox{red!40}{are} \colorbox{red!0}{collected} \colorbox{red!0}{from} \colorbox{red!0}{government} \colorbox{red!40}{agencies} \colorbox{red!20}{and} \colorbox{red!0}{companies} \colorbox{red!20}{.} \colorbox{red!0}{each} \colorbox{red!0}{salary} \colorbox{red!40}{is} \colorbox{red!0}{associated} \colorbox{red!20}{with} \colorbox{red!0}{a} \colorbox{red!0}{real} \colorbox{red!20}{job} \colorbox{red!40}{position} \colorbox{red!20}{.} \\
  \bottomrule
\end{tabular}
\end{table}

\section{Discussion}\label{sec:discussion}

In this section, we will discuss the results presented in Section~\ref{sec:results}, expounding upon several key discoveries that address the research questions proposed in Section~\ref{sec:intro}.

Regarding \textbf{RQ1} (\emph{How do the late-interaction models work and perform matching at the token level?}), we conducted experiments to explore the matching mechanisms of the late-interaction models, specifically to explore what document tokens will be selected and how
much they contribute to the final relevance score. Our findings indicate that both soft-matching models such as ColBERT, and hard-matching models such as COIL, heavily rely on important tokens for computing relevance scores. Specifically, tokens that are positioned towards the beginning of a paragraph or have a higher IDF value contribute more significantly to the calculation of the final relevance score. 

Inspired by the similarity between late-interaction models and traditional retrieval models, we propose to prune document tokens in order to achieve better efficiency while minimizing the loss of effectiveness. We conduct comprehensive experiments to examine the performance of various pruning techniques when employed on distinct models (\textbf{RQ2} and \textbf{RQ3}).

Regarding \textbf{RQ2} (\emph{What are the differences in performance among various models when applying pruning?}), we observe that the hard-matching late-interaction models, i.e., COIL and COIL-tok, are the most robust models when pruned on the supervised dataset. They can consistently achieve acceptable retrieval performance with different pruning methods and remaining ratios. The soft-matching late-interaction models, i.e., ColBERT and ColBERTv2, come next in terms of stability, while the sparse retrieval model is the least stable. We suggest this is because the hard-matching late-interaction models leverage the strengths of both dense and sparse retrieval techniques by utilizing the dense representations and exact match to compute the relevance score, thus achieving a better robustness when applying pruning although their original effectiveness is worse than the soft-matching late-interaction models. However, for experiments in the zero-shot setting, all models exhibited a decline in robustness to token pruning.

Regarding \textbf{RQ3} (\emph{How do different pruning methods and pruning ratios affect the effectiveness and efficiency of retrieval models?}), when we employ DTP and set the remaining ratio to a higher number, e.g., 50\% or 75\%, the First pruning is a simple but effective method across various models, the IDF-Top pruning shows unstable performance on different models, and the Attention-Top pruning method shows promising robustness on the supervised dataset but performs poorly on the zero-shot settings. This means that we can achieve a good trade-off between effectiveness and efficiency by pruning a small proportion of unimportant tokens. However, for all three methods, as the remaining ratio decreases to a smaller value, specifically 25\%, the retrieval performance has a significant loss, which might be unacceptable. Further, the QTP methods can reduce the retrieval latency with little loss of performance for ColBERT.

\section{Conclusion}\label{sec:conclusion}

This paper studies the late-interaction models when we employ different document token pruning methods. Firstly, we analyze late interaction models (i.e., ColBERT and COIL) from more traditional perspectives such as co-occurrence matches and term importance. Through the analysis, we further reveal the similarity between traditional models and late interaction models. Therefore, we are inspired by the match pattern of the traditional information retrieval models and study whether we can safely prune some tokens that are similar to the stop words in traditional retrieval models and which pruning method can achieve the best trade-off between effectiveness and efficiency. Results on the MS MARCO and BEIR show that we can reduce storage overhead with little performance loss when we use these late-interaction models. We further investigate the inference time, and query token pruning methods which skip some unnecessary ANNs with query tokens. Experimental results on the MS MARCO show that the query pruning method can speed up the retrieval with little loss of performance.

We acknowledge some potential limitations of this work. First of all, our proposed pruning methods are based on some simple yet effective heuristics, such as retaining the tokens with high IDF values or high attention sore. Therefore, they may not be the optimal pruning method as Attention-Top and IDF-Top exhibit different distributions on tokens. Secondly, when we conduct experiments on sparse models, we find that these pruning methods may not be as effective as directly removing lower-weighted tokens from the document. This indicates that the pruning strategies of sparse retrieval models could be rather different from those of typical late-interaction models with dense representations. We leave a more thorough investigation of the matching mechanism and effective pruning strategies for the sparse retrieval models to future work as they are out of the scope of this paper.

\begin{acks}
This research was supported by the Natural Science Foundation of China (61902209, 62377044, U2001212), and Beijing Outstanding Young Scientist Program (NO. JJWZYJH012019100020098), Intelligent Social Governance Platform, Major Innovation Planning Interdisciplinary Platform for the "Double-First Class" Initiative, Renmin University of China", the Fundamental Research Funds for the Central Universities, and the Research Funds of Renmin University of China.
\end{acks}

\bibliographystyle{ACM-Reference-Format}
\bibliography{references}


\begin{thebibliography}{50}


\ifx \showCODEN    \undefined \def \showCODEN     #1{\unskip}     \fi
\ifx \showDOI      \undefined \def \showDOI       #1{#1}\fi
\ifx \showISBNx    \undefined \def \showISBNx     #1{\unskip}     \fi
\ifx \showISBNxiii \undefined \def \showISBNxiii  #1{\unskip}     \fi
\ifx \showISSN     \undefined \def \showISSN      #1{\unskip}     \fi
\ifx \showLCCN     \undefined \def \showLCCN      #1{\unskip}     \fi
\ifx \shownote     \undefined \def \shownote      #1{#1}          \fi
\ifx \showarticletitle \undefined \def \showarticletitle #1{#1}   \fi
\ifx \showURL      \undefined \def \showURL       {\relax}        \fi
\providecommand\bibfield[2]{#2}
\providecommand\bibinfo[2]{#2}
\providecommand\natexlab[1]{#1}
\providecommand\showeprint[2][]{arXiv:#2}

\bibitem[Altin et~al\mbox{.}(2020)]%
        {altin2020pre}
\bibfield{author}{\bibinfo{person}{Soner Altin}, \bibinfo{person}{Ricardo
  Baeza-Yates}, {and} \bibinfo{person}{B~Barla Cambazoglu}.}
  \bibinfo{year}{2020}\natexlab{}.
\newblock \showarticletitle{Pre-indexing pruning strategies}. In
  \bibinfo{booktitle}{\emph{International Symposium on String Processing and
  Information Retrieval}}. Springer, \bibinfo{pages}{177--193}.
\newblock


\bibitem[Altingovde et~al\mbox{.}(2012)]%
        {altingovde2012static}
\bibfield{author}{\bibinfo{person}{Ismail~S Altingovde}, \bibinfo{person}{Rifat
  Ozcan}, {and} \bibinfo{person}{{\"O}zg{\"u}r Ulusoy}.}
  \bibinfo{year}{2012}\natexlab{}.
\newblock \showarticletitle{Static index pruning in web search engines:
  Combining term and document popularities with query views}.
\newblock \bibinfo{journal}{\emph{ACM Transactions on Information Systems
  (TOIS)}} \bibinfo{volume}{30}, \bibinfo{number}{1} (\bibinfo{year}{2012}),
  \bibinfo{pages}{1--28}.
\newblock


\bibitem[Baeza-Yates et~al\mbox{.}(2008)]%
        {baeza2008design}
\bibfield{author}{\bibinfo{person}{Ricardo Baeza-Yates},
  \bibinfo{person}{Aristides Gionis}, \bibinfo{person}{Flavio~P Junqueira},
  \bibinfo{person}{Vanessa Murdock}, \bibinfo{person}{Vassilis Plachouras},
  {and} \bibinfo{person}{Fabrizio Silvestri}.} \bibinfo{year}{2008}\natexlab{}.
\newblock \showarticletitle{Design trade-offs for search engine caching}.
\newblock \bibinfo{journal}{\emph{ACM Transactions on the Web (TWEB)}}
  \bibinfo{volume}{2}, \bibinfo{number}{4} (\bibinfo{year}{2008}),
  \bibinfo{pages}{1--28}.
\newblock


\bibitem[Bai et~al\mbox{.}(2020)]%
        {bai2020sparterm}
\bibfield{author}{\bibinfo{person}{Yang Bai}, \bibinfo{person}{Xiaoguang Li},
  \bibinfo{person}{Gang Wang}, \bibinfo{person}{Chaoliang Zhang},
  \bibinfo{person}{Lifeng Shang}, \bibinfo{person}{Jun Xu},
  \bibinfo{person}{Zhaowei Wang}, \bibinfo{person}{Fangshan Wang}, {and}
  \bibinfo{person}{Qun Liu}.} \bibinfo{year}{2020}\natexlab{}.
\newblock \showarticletitle{SparTerm: Learning Term-based Sparse Representation
  for Fast Text Retrieval}.
\newblock \bibinfo{journal}{\emph{arXiv preprint arXiv:2010.00768}}
  (\bibinfo{year}{2020}).
\newblock


\bibitem[Bajaj et~al\mbox{.}(2018)]%
        {bajaj2016msmarco}
\bibfield{author}{\bibinfo{person}{Payal Bajaj}, \bibinfo{person}{Daniel
  Campos}, \bibinfo{person}{Nick Craswell}, \bibinfo{person}{Li Deng},
  \bibinfo{person}{Jianfeng Gao}, \bibinfo{person}{Xiaodong Liu},
  \bibinfo{person}{Rangan Majumder}, \bibinfo{person}{Andrew McNamara},
  \bibinfo{person}{Bhaskar Mitra}, \bibinfo{person}{Tri Nguyen},
  \bibinfo{person}{Mir Rosenberg}, \bibinfo{person}{Xia Song},
  \bibinfo{person}{Alina Stoica}, \bibinfo{person}{Saurabh Tiwary}, {and}
  \bibinfo{person}{Tong Wang}.} \bibinfo{year}{2018}\natexlab{}.
\newblock \showarticletitle{MS MARCO: A Human Generated MAchine Reading
  COmprehension Dataset}.
\newblock \bibinfo{journal}{\emph{arXiv preprint arXiv:1611.09268}}
  (\bibinfo{year}{2018}).
\newblock


\bibitem[Blanco and Barreiro(2007)]%
        {blanco2007static}
\bibfield{author}{\bibinfo{person}{Roi Blanco} {and} \bibinfo{person}{Alvaro
  Barreiro}.} \bibinfo{year}{2007}\natexlab{}.
\newblock \showarticletitle{Static pruning of terms in inverted files}. In
  \bibinfo{booktitle}{\emph{Advances in Information Retrieval: 29th European
  Conference on IR Research, ECIR 2007, Rome, Italy, April 2-5, 2007.
  Proceedings 29}}. Springer, \bibinfo{pages}{64--75}.
\newblock


\bibitem[Blanco and Barreiro(2010)]%
        {blanco2010probabilistic}
\bibfield{author}{\bibinfo{person}{Roi Blanco} {and} \bibinfo{person}{Alvaro
  Barreiro}.} \bibinfo{year}{2010}\natexlab{}.
\newblock \showarticletitle{Probabilistic static pruning of inverted files}.
\newblock \bibinfo{journal}{\emph{ACM Transactions on Information Systems
  (TOIS)}} \bibinfo{volume}{28}, \bibinfo{number}{1} (\bibinfo{year}{2010}),
  \bibinfo{pages}{1--33}.
\newblock


\bibitem[B{\"u}ttcher and Clarke(2006)]%
        {buttcher2006document}
\bibfield{author}{\bibinfo{person}{Stefan B{\"u}ttcher} {and}
  \bibinfo{person}{Charles~LA Clarke}.} \bibinfo{year}{2006}\natexlab{}.
\newblock \showarticletitle{A document-centric approach to static index pruning
  in text retrieval systems}. In \bibinfo{booktitle}{\emph{Proceedings of the
  15th ACM international conference on Information and knowledge management}}.
  \bibinfo{pages}{182--189}.
\newblock


\bibitem[Carmel et~al\mbox{.}(2001)]%
        {carmel2001static}
\bibfield{author}{\bibinfo{person}{David Carmel}, \bibinfo{person}{Doron
  Cohen}, \bibinfo{person}{Ronald Fagin}, \bibinfo{person}{Eitan Farchi},
  \bibinfo{person}{Michael Herscovici}, \bibinfo{person}{Yoelle~S Maarek},
  {and} \bibinfo{person}{Aya Soffer}.} \bibinfo{year}{2001}\natexlab{}.
\newblock \showarticletitle{Static index pruning for information retrieval
  systems}. In \bibinfo{booktitle}{\emph{Proceedings of the 24th annual
  international ACM SIGIR conference on Research and development in information
  retrieval}}. \bibinfo{pages}{43--50}.
\newblock


\bibitem[Dai and Callan(2019)]%
        {dai2019context}
\bibfield{author}{\bibinfo{person}{Zhuyun Dai} {and} \bibinfo{person}{Jamie
  Callan}.} \bibinfo{year}{2019}\natexlab{}.
\newblock \showarticletitle{Context-aware sentence/passage term importance
  estimation for first stage retrieval}.
\newblock \bibinfo{journal}{\emph{arXiv preprint arXiv:1910.10687}}
  (\bibinfo{year}{2019}).
\newblock


\bibitem[Dai and Callan(2020)]%
        {dai2020context}
\bibfield{author}{\bibinfo{person}{Zhuyun Dai} {and} \bibinfo{person}{Jamie
  Callan}.} \bibinfo{year}{2020}\natexlab{}.
\newblock \showarticletitle{Context-aware document term weighting for ad-hoc
  search}. In \bibinfo{booktitle}{\emph{Proceedings of The Web Conference
  2020}}. \bibinfo{pages}{1897--1907}.
\newblock


\bibitem[Devlin et~al\mbox{.}(2018)]%
        {devlin2018bert}
\bibfield{author}{\bibinfo{person}{Jacob Devlin}, \bibinfo{person}{Ming-Wei
  Chang}, \bibinfo{person}{Kenton Lee}, {and} \bibinfo{person}{Kristina
  Toutanova}.} \bibinfo{year}{2018}\natexlab{}.
\newblock \showarticletitle{Bert: Pre-training of deep bidirectional
  transformers for language understanding}.
\newblock \bibinfo{journal}{\emph{arXiv preprint arXiv:1810.04805}}
  (\bibinfo{year}{2018}).
\newblock


\bibitem[Formal et~al\mbox{.}(2021a)]%
        {formal2021spladev2}
\bibfield{author}{\bibinfo{person}{Thibault Formal}, \bibinfo{person}{Carlos
  Lassance}, \bibinfo{person}{Benjamin Piwowarski}, {and}
  \bibinfo{person}{St{\'e}phane Clinchant}.} \bibinfo{year}{2021}\natexlab{a}.
\newblock \showarticletitle{SPLADE v2: Sparse lexical and expansion model for
  information retrieval}.
\newblock \bibinfo{journal}{\emph{arXiv preprint arXiv:2109.10086}}
  (\bibinfo{year}{2021}).
\newblock


\bibitem[Formal et~al\mbox{.}(2021b)]%
        {formal2021splade}
\bibfield{author}{\bibinfo{person}{Thibault Formal}, \bibinfo{person}{Benjamin
  Piwowarski}, {and} \bibinfo{person}{St{\'e}phane Clinchant}.}
  \bibinfo{year}{2021}\natexlab{b}.
\newblock \showarticletitle{SPLADE: Sparse lexical and expansion model for
  first stage ranking}. In \bibinfo{booktitle}{\emph{Proceedings of the 44th
  International ACM SIGIR Conference on Research and Development in Information
  Retrieval}}. \bibinfo{pages}{2288--2292}.
\newblock


\bibitem[Formal et~al\mbox{.}(2021c)]%
        {formal2021white}
\bibfield{author}{\bibinfo{person}{Thibault Formal}, \bibinfo{person}{Benjamin
  Piwowarski}, {and} \bibinfo{person}{St{\'e}phane Clinchant}.}
  \bibinfo{year}{2021}\natexlab{c}.
\newblock \showarticletitle{A white box analysis of colbert}. In
  \bibinfo{booktitle}{\emph{European Conference on Information Retrieval}}.
  Springer, \bibinfo{pages}{257--263}.
\newblock


\bibitem[Furnas et~al\mbox{.}(1987)]%
        {furnas1987vocabulary}
\bibfield{author}{\bibinfo{person}{George~W. Furnas},
  \bibinfo{person}{Thomas~K. Landauer}, \bibinfo{person}{Louis~M. Gomez}, {and}
  \bibinfo{person}{Susan~T. Dumais}.} \bibinfo{year}{1987}\natexlab{}.
\newblock \showarticletitle{The vocabulary problem in human-system
  communication}.
\newblock \bibinfo{journal}{\emph{Commun. ACM}} \bibinfo{volume}{30},
  \bibinfo{number}{11} (\bibinfo{year}{1987}), \bibinfo{pages}{964--971}.
\newblock


\bibitem[Gao et~al\mbox{.}(2021)]%
        {gao2021coil}
\bibfield{author}{\bibinfo{person}{Luyu Gao}, \bibinfo{person}{Zhuyun Dai},
  {and} \bibinfo{person}{Jamie Callan}.} \bibinfo{year}{2021}\natexlab{}.
\newblock \showarticletitle{COIL: Revisit Exact Lexical Match in Information
  Retrieval with Contextualized Inverted List}. In
  \bibinfo{booktitle}{\emph{Proceedings of the 2021 Conference of the North
  American Chapter of the Association for Computational Linguistics: Human
  Language Technologies}}. \bibinfo{pages}{3030--3042}.
\newblock


\bibitem[Hofst{\"a}tter et~al\mbox{.}(2022)]%
        {hofstatter2022introducing}
\bibfield{author}{\bibinfo{person}{Sebastian Hofst{\"a}tter},
  \bibinfo{person}{Omar Khattab}, \bibinfo{person}{Sophia Althammer},
  \bibinfo{person}{Mete Sertkan}, {and} \bibinfo{person}{Allan Hanbury}.}
  \bibinfo{year}{2022}\natexlab{}.
\newblock \showarticletitle{Introducing neural bag of whole-words with
  colberter: Contextualized late interactions using enhanced reduction}. In
  \bibinfo{booktitle}{\emph{Proceedings of the 31st ACM International
  Conference on Information \& Knowledge Management}}.
  \bibinfo{pages}{737--747}.
\newblock


\bibitem[Huang et~al\mbox{.}(2020)]%
        {huang2020embedding}
\bibfield{author}{\bibinfo{person}{Jui-Ting Huang}, \bibinfo{person}{Ashish
  Sharma}, \bibinfo{person}{Shuying Sun}, \bibinfo{person}{Li Xia},
  \bibinfo{person}{David Zhang}, \bibinfo{person}{Philip Pronin},
  \bibinfo{person}{Janani Padmanabhan}, \bibinfo{person}{Giuseppe Ottaviano},
  {and} \bibinfo{person}{Linjun Yang}.} \bibinfo{year}{2020}\natexlab{}.
\newblock \showarticletitle{Embedding-based retrieval in facebook search}. In
  \bibinfo{booktitle}{\emph{Proceedings of the 26th ACM SIGKDD International
  Conference on Knowledge Discovery \& Data Mining}}.
  \bibinfo{pages}{2553--2561}.
\newblock


\bibitem[Humeau et~al\mbox{.}(2019)]%
        {humeau2019poly}
\bibfield{author}{\bibinfo{person}{Samuel Humeau}, \bibinfo{person}{Kurt
  Shuster}, \bibinfo{person}{Marie-Anne Lachaux}, {and} \bibinfo{person}{Jason
  Weston}.} \bibinfo{year}{2019}\natexlab{}.
\newblock \showarticletitle{Poly-encoders: Architectures and Pre-training
  Strategies for Fast and Accurate Multi-sentence Scoring}. In
  \bibinfo{booktitle}{\emph{International Conference on Learning
  Representations}}.
\newblock


\bibitem[Johnson et~al\mbox{.}(2019)]%
        {johnson2019billion}
\bibfield{author}{\bibinfo{person}{Jeff Johnson}, \bibinfo{person}{Matthijs
  Douze}, {and} \bibinfo{person}{Herv{\'e} J{\'e}gou}.}
  \bibinfo{year}{2019}\natexlab{}.
\newblock \showarticletitle{Billion-scale similarity search with gpus}.
\newblock \bibinfo{journal}{\emph{IEEE Transactions on Big Data}}
  (\bibinfo{year}{2019}).
\newblock


\bibitem[Karpukhin et~al\mbox{.}(2020)]%
        {karpukhin2020dense}
\bibfield{author}{\bibinfo{person}{Vladimir Karpukhin}, \bibinfo{person}{Barlas
  Oguz}, \bibinfo{person}{Sewon Min}, \bibinfo{person}{Patrick Lewis},
  \bibinfo{person}{Ledell Wu}, \bibinfo{person}{Sergey Edunov},
  \bibinfo{person}{Danqi Chen}, {and} \bibinfo{person}{Wen-tau Yih}.}
  \bibinfo{year}{2020}\natexlab{}.
\newblock \showarticletitle{Dense Passage Retrieval for Open-Domain Question
  Answering}. In \bibinfo{booktitle}{\emph{Proceedings of the 2020 Conference
  on Empirical Methods in Natural Language Processing (EMNLP)}}.
  \bibinfo{pages}{6769--6781}.
\newblock


\bibitem[Khattab and Zaharia(2020)]%
        {khattab2020colbert}
\bibfield{author}{\bibinfo{person}{Omar Khattab} {and} \bibinfo{person}{Matei
  Zaharia}.} \bibinfo{year}{2020}\natexlab{}.
\newblock \showarticletitle{Colbert: Efficient and effective passage search via
  contextualized late interaction over bert}. In
  \bibinfo{booktitle}{\emph{Proceedings of the 43rd International ACM SIGIR
  conference on research and development in Information Retrieval}}.
  \bibinfo{pages}{39--48}.
\newblock


\bibitem[Lan et~al\mbox{.}(2019)]%
        {lan2019albert}
\bibfield{author}{\bibinfo{person}{Zhenzhong Lan}, \bibinfo{person}{Mingda
  Chen}, \bibinfo{person}{Sebastian Goodman}, \bibinfo{person}{Kevin Gimpel},
  \bibinfo{person}{Piyush Sharma}, {and} \bibinfo{person}{Radu Soricut}.}
  \bibinfo{year}{2019}\natexlab{}.
\newblock \showarticletitle{ALBERT: A Lite BERT for Self-supervised Learning of
  Language Representations}. In \bibinfo{booktitle}{\emph{International
  Conference on Learning Representations}}.
\newblock


\bibitem[Lassance et~al\mbox{.}(2023)]%
        {lassance2023static}
\bibfield{author}{\bibinfo{person}{Carlos Lassance}, \bibinfo{person}{Simon
  Lupart}, \bibinfo{person}{Herv{\'e} Dejean}, \bibinfo{person}{St{\'e}phane
  Clinchant}, {and} \bibinfo{person}{Nicola Tonellotto}.}
  \bibinfo{year}{2023}\natexlab{}.
\newblock \showarticletitle{A Static Pruning Study on Sparse Neural
  Retrievers}.
\newblock \bibinfo{journal}{\emph{arXiv preprint arXiv:2304.12702}}
  (\bibinfo{year}{2023}).
\newblock


\bibitem[Lassance et~al\mbox{.}(2021)]%
        {lassance2021study}
\bibfield{author}{\bibinfo{person}{Carlos Lassance}, \bibinfo{person}{Maroua
  Maachou}, \bibinfo{person}{Joohee Park}, {and} \bibinfo{person}{St{\'e}phane
  Clinchant}.} \bibinfo{year}{2021}\natexlab{}.
\newblock \showarticletitle{A Study on Token Pruning for ColBERT}.
\newblock \bibinfo{journal}{\emph{arXiv preprint arXiv:2112.06540}}
  (\bibinfo{year}{2021}).
\newblock


\bibitem[Lin and Ma(2021)]%
        {lin2021few}
\bibfield{author}{\bibinfo{person}{Jimmy Lin} {and} \bibinfo{person}{Xueguang
  Ma}.} \bibinfo{year}{2021}\natexlab{}.
\newblock \showarticletitle{A few brief notes on deepimpact, coil, and a
  conceptual framework for information retrieval techniques}.
\newblock \bibinfo{journal}{\emph{arXiv preprint arXiv:2106.14807}}
  (\bibinfo{year}{2021}).
\newblock


\bibitem[Lin et~al\mbox{.}(2021)]%
        {lin2021pyserini}
\bibfield{author}{\bibinfo{person}{Jimmy Lin}, \bibinfo{person}{Xueguang Ma},
  \bibinfo{person}{Sheng-Chieh Lin}, \bibinfo{person}{Jheng-Hong Yang},
  \bibinfo{person}{Ronak Pradeep}, {and} \bibinfo{person}{Rodrigo Nogueira}.}
  \bibinfo{year}{2021}\natexlab{}.
\newblock \showarticletitle{{Pyserini}: A {Python} Toolkit for Reproducible
  Information Retrieval Research with Sparse and Dense Representations}. In
  \bibinfo{booktitle}{\emph{Proceedings of the 44th Annual International ACM
  SIGIR Conference on Research and Development in Information Retrieval (SIGIR
  2021)}}. \bibinfo{pages}{2356--2362}.
\newblock


\bibitem[Liu et~al\mbox{.}(2019)]%
        {liu2019roberta}
\bibfield{author}{\bibinfo{person}{Yinhan Liu}, \bibinfo{person}{Myle Ott},
  \bibinfo{person}{Naman Goyal}, \bibinfo{person}{Jingfei Du},
  \bibinfo{person}{Mandar Joshi}, \bibinfo{person}{Danqi Chen},
  \bibinfo{person}{Omer Levy}, \bibinfo{person}{Mike Lewis},
  \bibinfo{person}{Luke Zettlemoyer}, {and} \bibinfo{person}{Veselin
  Stoyanov}.} \bibinfo{year}{2019}\natexlab{}.
\newblock \showarticletitle{Roberta: A robustly optimized bert pretraining
  approach}.
\newblock \bibinfo{journal}{\emph{arXiv preprint arXiv:1907.11692}}
  (\bibinfo{year}{2019}).
\newblock


\bibitem[Luan et~al\mbox{.}(2021)]%
        {luan2020sparse}
\bibfield{author}{\bibinfo{person}{Yi Luan}, \bibinfo{person}{Jacob
  Eisenstein}, \bibinfo{person}{Kristina Toutanova}, {and}
  \bibinfo{person}{Michael Collins}.} \bibinfo{year}{2021}\natexlab{}.
\newblock \showarticletitle{Sparse, Dense, and Attentional Representations for
  Text Retrieval}.
\newblock \bibinfo{journal}{\emph{Transactions of the Association for
  Computational Linguistics}}  \bibinfo{volume}{9} (\bibinfo{year}{2021}),
  \bibinfo{pages}{329--345}.
\newblock


\bibitem[Mallia et~al\mbox{.}(2021)]%
        {mallia2021learning}
\bibfield{author}{\bibinfo{person}{Antonio Mallia}, \bibinfo{person}{Omar
  Khattab}, \bibinfo{person}{Torsten Suel}, {and} \bibinfo{person}{Nicola
  Tonellotto}.} \bibinfo{year}{2021}\natexlab{}.
\newblock \showarticletitle{Learning passage impacts for inverted indexes}. In
  \bibinfo{booktitle}{\emph{Proceedings of the 44th International ACM SIGIR
  Conference on Research and Development in Information Retrieval}}.
  \bibinfo{pages}{1723--1727}.
\newblock


\bibitem[Nogueira and Cho(2019)]%
        {nogueira2019passage}
\bibfield{author}{\bibinfo{person}{Rodrigo Nogueira} {and}
  \bibinfo{person}{Kyunghyun Cho}.} \bibinfo{year}{2019}\natexlab{}.
\newblock \showarticletitle{Passage Re-ranking with BERT}.
\newblock \bibinfo{journal}{\emph{arXiv preprint arXiv:1901.04085}}
  (\bibinfo{year}{2019}).
\newblock


\bibitem[Nogueira et~al\mbox{.}(2019a)]%
        {nogueira2019doc2query}
\bibfield{author}{\bibinfo{person}{Rodrigo Nogueira}, \bibinfo{person}{Jimmy
  Lin}, {and} \bibinfo{person}{AI Epistemic}.}
  \bibinfo{year}{2019}\natexlab{a}.
\newblock \showarticletitle{From doc2query to docTTTTTquery}.
\newblock \bibinfo{journal}{\emph{Online preprint}} (\bibinfo{year}{2019}).
\newblock


\bibitem[Nogueira et~al\mbox{.}(2019b)]%
        {nogueira2019document}
\bibfield{author}{\bibinfo{person}{Rodrigo Nogueira}, \bibinfo{person}{Wei
  Yang}, \bibinfo{person}{Jimmy Lin}, {and} \bibinfo{person}{Kyunghyun Cho}.}
  \bibinfo{year}{2019}\natexlab{b}.
\newblock \showarticletitle{Document expansion by query prediction}.
\newblock \bibinfo{journal}{\emph{arXiv preprint arXiv:1904.08375}}
  (\bibinfo{year}{2019}).
\newblock


\bibitem[Ponte and Croft(1998)]%
        {ponte1998language}
\bibfield{author}{\bibinfo{person}{Jay~M Ponte} {and} \bibinfo{person}{W~Bruce
  Croft}.} \bibinfo{year}{1998}\natexlab{}.
\newblock \showarticletitle{A language modeling approach to information
  retrieval}. In \bibinfo{booktitle}{\emph{Proceedings of the 21st annual
  international ACM SIGIR conference on Research and development in information
  retrieval}}. \bibinfo{pages}{275--281}.
\newblock


\bibitem[Qian et~al\mbox{.}(2022)]%
        {qian2022multi}
\bibfield{author}{\bibinfo{person}{Yujie Qian}, \bibinfo{person}{Jinhyuk Lee},
  \bibinfo{person}{Sai Meher~Karthik Duddu}, \bibinfo{person}{Zhuyun Dai},
  \bibinfo{person}{Siddhartha Brahma}, \bibinfo{person}{Iftekhar Naim},
  \bibinfo{person}{Tao Lei}, {and} \bibinfo{person}{Vincent~Y Zhao}.}
  \bibinfo{year}{2022}\natexlab{}.
\newblock \showarticletitle{Multi-vector retrieval as sparse alignment}.
\newblock \bibinfo{journal}{\emph{arXiv preprint arXiv:2211.01267}}
  (\bibinfo{year}{2022}).
\newblock


\bibitem[Qiao et~al\mbox{.}(2019)]%
        {qiao2019understanding}
\bibfield{author}{\bibinfo{person}{Yifan Qiao}, \bibinfo{person}{Chenyan
  Xiong}, \bibinfo{person}{Zhenghao Liu}, {and} \bibinfo{person}{Zhiyuan Liu}.}
  \bibinfo{year}{2019}\natexlab{}.
\newblock \showarticletitle{Understanding the Behaviors of BERT in Ranking}.
\newblock \bibinfo{journal}{\emph{arXiv preprint arXiv:1904.07531}}
  (\bibinfo{year}{2019}).
\newblock


\bibitem[Radford et~al\mbox{.}(2019)]%
        {radford2019language}
\bibfield{author}{\bibinfo{person}{Alec Radford}, \bibinfo{person}{Jeffrey Wu},
  \bibinfo{person}{Rewon Child}, \bibinfo{person}{David Luan},
  \bibinfo{person}{Dario Amodei}, {and} \bibinfo{person}{Ilya Sutskever}.}
  \bibinfo{year}{2019}\natexlab{}.
\newblock \showarticletitle{Language models are unsupervised multitask
  learners}.
\newblock \bibinfo{journal}{\emph{OpenAI blog}} \bibinfo{volume}{1},
  \bibinfo{number}{8} (\bibinfo{year}{2019}), \bibinfo{pages}{9}.
\newblock


\bibitem[Raffel et~al\mbox{.}(2019)]%
        {raffel2019exploring}
\bibfield{author}{\bibinfo{person}{Colin Raffel}, \bibinfo{person}{Noam
  Shazeer}, \bibinfo{person}{Adam Roberts}, \bibinfo{person}{Katherine Lee},
  \bibinfo{person}{Sharan Narang}, \bibinfo{person}{Michael Matena},
  \bibinfo{person}{Yanqi Zhou}, \bibinfo{person}{Wei Li}, {and}
  \bibinfo{person}{Peter~J Liu}.} \bibinfo{year}{2019}\natexlab{}.
\newblock \showarticletitle{Exploring the limits of transfer learning with a
  unified text-to-text transformer}.
\newblock \bibinfo{journal}{\emph{arXiv preprint arXiv:1910.10683}}
  (\bibinfo{year}{2019}).
\newblock


\bibitem[Robertson et~al\mbox{.}(2009)]%
        {robertson2009probabilistic}
\bibfield{author}{\bibinfo{person}{Stephen Robertson}, \bibinfo{person}{Hugo
  Zaragoza}, {et~al\mbox{.}}} \bibinfo{year}{2009}\natexlab{}.
\newblock \showarticletitle{The probabilistic relevance framework: BM25 and
  beyond}.
\newblock \bibinfo{journal}{\emph{Foundations and Trends{\textregistered} in
  Information Retrieval}} \bibinfo{volume}{3}, \bibinfo{number}{4}
  (\bibinfo{year}{2009}), \bibinfo{pages}{333--389}.
\newblock


\bibitem[Santhanam et~al\mbox{.}(2022)]%
        {santhanam2022colbertv2}
\bibfield{author}{\bibinfo{person}{Keshav Santhanam}, \bibinfo{person}{Omar
  Khattab}, \bibinfo{person}{Jon Saad-Falcon}, \bibinfo{person}{Christopher
  Potts}, {and} \bibinfo{person}{Matei Zaharia}.}
  \bibinfo{year}{2022}\natexlab{}.
\newblock \showarticletitle{ColBERTv2: Effective and Efficient Retrieval via
  Lightweight Late Interaction}. In \bibinfo{booktitle}{\emph{Proceedings of
  the 2022 Conference of the North American Chapter of the Association for
  Computational Linguistics: Human Language Technologies}}.
  \bibinfo{pages}{3715--3734}.
\newblock


\bibitem[Schuirmann(1987)]%
        {schuirmann1987tost}
\bibfield{author}{\bibinfo{person}{Donald~J Schuirmann}.}
  \bibinfo{year}{1987}\natexlab{}.
\newblock \showarticletitle{A comparison of the two one-sided tests procedure
  and the power approach for assessing the equivalence of average
  bioavailability}.
\newblock \bibinfo{journal}{\emph{Journal of pharmacokinetics and
  biopharmaceutics}}  \bibinfo{volume}{15} (\bibinfo{year}{1987}),
  \bibinfo{pages}{657--680}.
\newblock


\bibitem[Thakur et~al\mbox{.}(2021)]%
        {thakur2021beir}
\bibfield{author}{\bibinfo{person}{Nandan Thakur}, \bibinfo{person}{Nils
  Reimers}, \bibinfo{person}{Andreas R{\"u}ckl{\'e}}, \bibinfo{person}{Abhishek
  Srivastava}, {and} \bibinfo{person}{Iryna Gurevych}.}
  \bibinfo{year}{2021}\natexlab{}.
\newblock \showarticletitle{BEIR: A Heterogeneous Benchmark for Zero-shot
  Evaluation of Information Retrieval Models}. In
  \bibinfo{booktitle}{\emph{Thirty-fifth Conference on Neural Information
  Processing Systems Datasets and Benchmarks Track (Round 2)}}.
\newblock


\bibitem[Thota and Carterette(2011)]%
        {thota2011within}
\bibfield{author}{\bibinfo{person}{Sree~Lekha Thota} {and} \bibinfo{person}{Ben
  Carterette}.} \bibinfo{year}{2011}\natexlab{}.
\newblock \showarticletitle{Within-document term-based index pruning with
  statistical hypothesis testing}. In \bibinfo{booktitle}{\emph{Advances in
  Information Retrieval: 33rd European Conference on IR Research, ECIR 2011,
  Dublin, Ireland, April 18-21, 2011. Proceedings 33}}. Springer,
  \bibinfo{pages}{543--554}.
\newblock


\bibitem[Tonellotto and Macdonald(2021)]%
        {tonellotto2021query}
\bibfield{author}{\bibinfo{person}{Nicola Tonellotto} {and}
  \bibinfo{person}{Craig Macdonald}.} \bibinfo{year}{2021}\natexlab{}.
\newblock \showarticletitle{Query embedding pruning for dense retrieval}. In
  \bibinfo{booktitle}{\emph{Proceedings of the 30th ACM International
  Conference on Information \& Knowledge Management}}.
  \bibinfo{pages}{3453--3457}.
\newblock


\bibitem[Xiong et~al\mbox{.}(2020)]%
        {xiong2020approximate}
\bibfield{author}{\bibinfo{person}{Lee Xiong}, \bibinfo{person}{Chenyan Xiong},
  \bibinfo{person}{Ye Li}, \bibinfo{person}{Kwok-Fung Tang},
  \bibinfo{person}{Jialin Liu}, \bibinfo{person}{Paul~N Bennett},
  \bibinfo{person}{Junaid Ahmed}, {and} \bibinfo{person}{Arnold Overwijk}.}
  \bibinfo{year}{2020}\natexlab{}.
\newblock \showarticletitle{Approximate Nearest Neighbor Negative Contrastive
  Learning for Dense Text Retrieval}. In
  \bibinfo{booktitle}{\emph{International Conference on Learning
  Representations}}.
\newblock


\bibitem[Yang et~al\mbox{.}(2019)]%
        {yang2019xlnet}
\bibfield{author}{\bibinfo{person}{Zhilin Yang}, \bibinfo{person}{Zihang Dai},
  \bibinfo{person}{Yiming Yang}, \bibinfo{person}{Jaime Carbonell},
  \bibinfo{person}{Russ~R Salakhutdinov}, {and} \bibinfo{person}{Quoc~V Le}.}
  \bibinfo{year}{2019}\natexlab{}.
\newblock \showarticletitle{Xlnet: Generalized autoregressive pretraining for
  language understanding}.
\newblock \bibinfo{journal}{\emph{Advances in neural information processing
  systems}}  \bibinfo{volume}{32} (\bibinfo{year}{2019}).
\newblock


\bibitem[Zhan et~al\mbox{.}(2021)]%
        {zhan2021optimizing}
\bibfield{author}{\bibinfo{person}{Jingtao Zhan}, \bibinfo{person}{Jiaxin Mao},
  \bibinfo{person}{Yiqun Liu}, \bibinfo{person}{Jiafeng Guo},
  \bibinfo{person}{Min Zhang}, {and} \bibinfo{person}{Shaoping Ma}.}
  \bibinfo{year}{2021}\natexlab{}.
\newblock \showarticletitle{Optimizing dense retrieval model training with hard
  negatives}. In \bibinfo{booktitle}{\emph{Proceedings of the 44th
  International ACM SIGIR Conference on Research and Development in Information
  Retrieval}}. \bibinfo{pages}{1503--1512}.
\newblock


\bibitem[Zhan et~al\mbox{.}(2020a)]%
        {zhan2020analysis}
\bibfield{author}{\bibinfo{person}{Jingtao Zhan}, \bibinfo{person}{Jiaxin Mao},
  \bibinfo{person}{Yiqun Liu}, \bibinfo{person}{Min Zhang}, {and}
  \bibinfo{person}{Shaoping Ma}.} \bibinfo{year}{2020}\natexlab{a}.
\newblock \showarticletitle{An analysis of BERT in document ranking}. In
  \bibinfo{booktitle}{\emph{Proceedings of the 43rd International ACM SIGIR
  conference on research and development in Information Retrieval}}.
  \bibinfo{pages}{1941--1944}.
\newblock


\bibitem[Zhan et~al\mbox{.}(2020b)]%
        {zhan2020repbert}
\bibfield{author}{\bibinfo{person}{Jingtao Zhan}, \bibinfo{person}{Jiaxin Mao},
  \bibinfo{person}{Yiqun Liu}, \bibinfo{person}{Min Zhang}, {and}
  \bibinfo{person}{Shaoping Ma}.} \bibinfo{year}{2020}\natexlab{b}.
\newblock \showarticletitle{RepBERT: Contextualized text embeddings for
  first-stage retrieval}.
\newblock \bibinfo{journal}{\emph{arXiv preprint arXiv:2006.15498}}
  (\bibinfo{year}{2020}).
\newblock


\end{thebibliography}

\end{document}